\documentclass[11pt]{article} 
\usepackage{jheppub}
\usepackage{amsfonts,ulem,bbold}
\usepackage{amsmath,amssymb, braket}
\newcommand{\te}{\tilde\epsilon}

\newcommand{\be}{\begin{equation}}
\newcommand{\ee}{\end{equation}}
\newcommand{\bea}{\begin{eqnarray}}
\newcommand{\eea}{\end{eqnarray}}
\newcommand{\beqn}{\begin{eqnarray}}
\newcommand{\eeqn}{\end{eqnarray}}
\newcommand{\p}{\partial}
\newcommand{\nn}{\nonumber}
\newcommand{\wt}{\widetilde}
\newcommand{\ds}{\displaystyle}
\newcommand{\gmn}{g_{\mu\nu}}
\newcommand{\fmn}{f_{\mu\nu}}
\newcommand{\Gmn}{G_{\mu\nu}}
\newcommand{\Mmn}{M_{\mu\nu}}
\newcommand{\bgmn}{\bar g_{\mu\nu}}
\newcommand{\bfmn}{\bar f_{\mu\nu}}

\newcommand{\meff}{M_\mathrm{eff}}

\newcommand{\dg}{\delta{g}}
\newcommand{\df}{\delta{f}}

\newcommand{\dS}{\delta{S}}

\newcommand{\dMg}{\delta M}

\newcommand{\bg}{\bar{g}}

\newcommand{\mfp}{m_{\mathrm{FP}}}

\def\Lag{\mathcal{L}}
\def\half{\frac{1}{2}}
\def\Lag{\mathcal{L}}

\def\meff{m_{\mathrm{eff}}}
\def\td{\mathrm{d}}
\newcommand{\md}{\mathrm{d}}
\def\ph{\phantom}

\title{On Consistent Theories of Massive Spin-2 Fields Coupled to Gravity}
\author{S.F.~Hassan,}
\author{Angnis~Schmidt-May}
\author{and Mikael~von~Strauss}

\affiliation{Department of Physics \& 
        The Oskar Klein Centre,\\
        Stockholm University, AlbaNova University Centre, 
        SE-106 91 Stockholm, Sweden}

\emailAdd{fawad@fysik.su.se}
\emailAdd{angnis.schmidt-may@fysik.su.se}
\emailAdd{mvs@fysik.su.se}

\abstract{We consider the issues that arise out of interpreting the
  ghost-free bimetric theory as a theory of a spin-2 field coupled to
  gravity. This requires identifying a gravitational metric and
  parameterizing deviations of the resulting theory from general
  relativity. To this end, we first consider the most general bimetric
  backgrounds for which a massless and a massive spin-2 fluctuation
  exist, and we compute the most general expression for the
  Fierz-Pauli mass. These backgrounds coincide with solutions in
  general relativity. Based on this, we obtain nonlinear extensions of
  the massive and massless spin-2 fields. The background value of the
  nonlinear massive field parameterizes generic deviations of the
  bimetric theory from GR. It is also shown that the most natural nonlinear
  massless field does not have standard ghost-free matter couplings,
  and hence cannot represent the gravitational metric. However, an
  appropriate gravitational metric can still be identified in the weak
  gravity limit. Hence in the presence of other neutral spin-2 fields,
  the weak gravity limit is crucial for compatibility with general
  relativity. We also write down the action in terms of the nonlinear
  massive spin-2 field and obtain its ghost-free couplings to matter.
  The discussion is then generalized to multimetric theories.}


\toccontinuoustrue

\begin{document} 
\maketitle
\flushbottom

\section{Introduction, motivation and summary}

Theories of interacting spin-2 fields have been considered over many
years with various motivations, for example, in \cite{Rosen:1940zz,
  ISS,Aragone:1971kh,Rosen:1975kk,Chamseddine:1978yu,Print-79-0993},
or more recently in \cite{hep-th/9908028,hep-th/9910188,
  Boulanger:2000rq,Damour:2002ws,Damour:2002wu,ArkaniHamed:2002sp,
  Blas:2007zz,Banados:2008fi,Milgrom:2009gv}. Often, these are
formulated in terms of two metrics, $\gmn$ and $\fmn$, with
non-derivative interactions. These theories generically contain
Boulware-Deser ghost instabilities \cite{Boulware:1973my}. The bimetric
theories that avoid this problem were written down and proven to be
ghost-free in \cite{Hassan:2011zd,arXiv:1111.2070}. This was based on
\cite{Hassan:2011vm,Hassan:2011hr,Hassan:2011tf} that further
developed the massive gravity work in
\cite{deRham:2010ik,deRham:2010kj}, as will be briefly reviewed
below. More recently, this was extended to ghost-free theories of many
spin-2 fields in terms of ${\mathcal N}$ vielbeins
\cite{Hinterbichler:2012cn}, while a formulation in terms of
${\mathcal N}$ metrics is given in \cite{Hassan:2012wt}.\footnote{In
  hindsight, it turns out that Chamseddine, Salam and Strathdee in
  1978 \cite{Chamseddine:1978yu} had a ghost-free bimetric theory,
  written in terms of vielbeins and with supersymmetry, although the
  absence of the BD ghost could not be demonstrated then.}

In the ghost-free bimetric theory \cite{Hassan:2011zd}, a priori, the
two spin-2 fields $\gmn$ and $\fmn$ appear more or less on the same
footing. For obvious reasons, eventually we would like to interpret
this as a theory of a ``massive'' spin-2 field interacting with
gravity. Furthermore, the gravity sector of the theory should not show
observable deviations from tested aspects of general relativity. In
this paper we consider the issues that arise when using the bimetric
theory to describe a spin-2 field coupled to gravity.  The
considerations also apply to the multivielbein/multimetric case.

To focus attention, the ghost-free bimetric theory that we will work
with has the form,   
\be
S=\int\td^4x\Big[m^2_g\sqrt{-g}\,R(g)+m^2_f\sqrt{-f}
\, R(f)- 2m^4 \sqrt{-g}\,\,V(g^{-1}f\,,\beta_n)\Big]\,
+S_m(g,f,\psi_m)\,, 
\label{action-gf_rev} 
\ee
with details to be specified later \eqref{action}, \eqref{int_pot}. The particular combination of
kinetic and potential terms renders the theory ghost-free. The seven 
parameters of the theory are $m_p$, $m_f$ and five $\beta_n$. The
simplest possible ``matter'' interactions that are also known to be  
ghost-free \cite{Hassan:2011zd}, are of the form, 
\be
\sqrt{-g}\,{\cal L}_g(g,\psi)+
\sqrt{-f}\,{\cal L}_f(f,\psi')\,.
\label{action-gf-m_rev} 
\ee
Other forms of matter coupling should be explicitly checked for ghosts.  
Generic cosmological and localized solutions in this theory could show
large deviations from solutions in general relativity (GR) although
there also exist classes of solutions that are close to GR spacetimes
\cite{Volkov:2011an,Volkov:2012wp,vonStrauss:2011mq,Comelli:2011wq, 
Comelli:2011zm,Baccetti:2012ge,Deffayet:2011rh}.   

Below, we will first describe the issues that arise out of interpreting
(\ref{action-gf_rev}) as a theory of a spin-2 field coupled to
gravity, and summarize our results. Then we will briefly review the
development of spin-2 theories with emphasis on the importance of the
nonlinear methods. 

\subsection{Issues considered and summary of results} 
\label{Need}

By construction, in (\ref{action-gf_rev}) around flat backgrounds
$\bgmn= \bfmn =\eta_{\mu\nu}$ (that exist for a restricted set of
$\beta_n$), the fluctuations $\delta\gmn$ and $\delta\fmn$ are linear
combinations of a massless spin-2 mode $\delta\Gmn$ (2 polarizations)
and a massive spin-2 mode $\delta\Mmn$ (5 polarizations) with a
Fierz-Pauli mass term \cite{Fierz:1939ix,Pauli:1939xp}. At nonlinear
level too, the theory has 7 propagating degrees of freedom, although
in that case the analogue of the decomposition in terms of mass is not
known. An obvious problem is to specify the most general class of
backgrounds around which the theory has well defined massive and
massless fluctuations, and to compute the spectrum as a function of
the unrestricted $\beta_n$. 

To regard (\ref{action-gf_rev}) as a theory of a neutral spin-2 field
interacting with gravity, one has to first identify the gravitational
metric, say $g^{\text{GR}}$, in terms of $g$ and $f$. An important
restriction is that the standard minimal couplings of $g^{\text{GR}}$ to
matter, demanded by the weak equivalence principle, should also be
ghost-free. A first guess for $g^{\text{GR}}$, one suggested as far back as
\cite{ISS}, is the nonlinear extension of the massless mode
$\delta\Gmn$. But to explicitly check if this allows for ghost-free
matter couplings, one needs an explicit nonlinear expression for it in
terms of $g$ and $f$. The other obvious fall back options are $g$ or
$f$. While not mass eigenstates, these have ghost-free matter
couplings. 

Having identified a gravitational metric, the next task is to verify
that the theory has parameter space regions where the solutions for 
$g^{\text{GR}}$ are close enough to GR solutions that the bimetric theory is
not immediately ruled out on observational grounds. For this, if
possible, one would like to have some criteria or quantity to
parameterize deviations of the bimetric theory from GR. In this paper
we consider these issues and the results are summarized below.

{\bf Proportional backgrounds and general mass eigenstates:} To obtain
the mass spectrum, we consider the most general class of bimetric
backgrounds around which a massive mode with a well defined
Fierz-Pauli mass term exists. These are the proportional backgrounds
$\bfmn=c^2\bgmn$, where $c$ is determined by the parameters of the
theory. They coincide exactly with solutions in general relativity
with a cosmological constant, and always exist as bimetric vacuum
solutions without fixing the parameters of the theory, as long as real 
solutions for $c$ exist. Flat space solutions require fixing one
of the seven parameters by setting the cosmological constant to zero.
The solutions also exist in the presence of sources, as long as the
sources of the $g$ and $f$ equations of motion satisfy $m_g^2\bar
T^f_{\mu\nu}=m_f^{2}\, \bar T^g_{\mu\nu}$. This constraint is not
natural, but shows that deviations from it drive bimetric solutions
away from GR solutions in a generic sense (although it is still
possible to get isolated GR type solutions).

Considering fluctuations around $\bfmn=c^2\bgmn$ backgrounds, we
obtain the most general expression for the Fierz-Pauli mass, as well
as the expressions for the massless mode $\delta\Gmn$
and the massive mode $\delta\Mmn$.

{\bf Nonlinear modes:} We give a procedure to systematically obtain
nonlinear combinations of $f$ and $g$ that reduce to $\delta G$ and
$\delta M$ at the linear level. Although there are infinitely many such combinations we identify one, $\Gmn$, as
the nonlinear extension of the massless mode and two possible
candidates, $\Mmn$ and $\Mmn^G$, for the nonlinear extension of the
massive mode, based on reasonable criteria. These seem natural and are
simple enough that the expressions relating them to $g$ and $f$ are
invertible. The vanishing of the nonlinear massive mode, $M=0$, is in
one-to-one correspondence with occurrence of proportional backgrounds
$\bfmn=c^2\bgmn$. Hence deviations of the VEV of $M$ from $0$ are
driven by the matter couplings of the spin-2 fields and parameterize
generic deviations of the bimetric theory from GR. If these nonlinear
modes have a relevance directly at the nonlinear level is not yet
answered.

{\bf Identification of gravity:} Having a nonlinear massless mode $G$
in hand, we can test the conjecture that it should be identified as
the gravitational metric. Through an ADM analysis we show that within
the bimetric framework, the standard minimal coupling of $\Gmn$ to
matter is not ghost-free. This rules out that particular $\Gmn$ as a candidate for the 
gravitational metric. 

Another option is $\gmn$ (or equivalently $\fmn$ since the formulation
is symmetric) which has a ghost-free matter coupling. In particular,
in the limit $m_g>> c\,m_f$, we have $\delta G \rightarrow \delta g$
and $\delta M\rightarrow\delta f$. For the nonlinear fields too, in
the limit, $m_g>>m_f$, $G \rightarrow g$, although in this case $M$
has no particular limit. Hence, if one identifies $\gmn$ as the
gravitational metric, then in the weak gravity limit, $\gmn$ will
mostly consist of the massless mode. An obvious consequence is that in
the presence of massive spin-2 fields, metric perturbations created by
a matter source will also have a small massive component.

Now one can regard the pair $g,M$ as the basic variables and express
the bimetric action in terms of them. Although the kinetic part in 
terms of $M$ is more involved than the original form in terms of $f$,
the potential is now a finite polynomial in $M$ and does not involve a
square-root matrix. Also as pointed out earlier, couplings that drive
$M$ away from $\bar M=0$, also drive the solutions for the metric $g$
away from GR. Subsequently, from the ghost-free couplings of $\fmn$ to
matter, we obtain couplings between the massive field $M$ and matter. 

{\bf Multi spin-2 fields coupled to gravity:} Finally we extend the
above considerations to multi spin-2 theories, as theories of
${\mathcal N}-1$ massive spin-2 fields coupled to gravity.

\subsection{Background to bimetric theories} 
The BD ghost was first observed in massive gravity
\cite{Boulware:1973my} which corresponds to the bimetric theory with
one metric held fixed, say, $\fmn=\eta_{\mu\nu}$. It led to the
speculation that such ghost-free theories may not exist. The major
breakthrough came with the work of de Rham, Gabadadze and Tolley
\cite{deRham:2010ik,deRham:2010kj}, who obtained a potentially
ghost-free nonlinear massive gravity action for $\fmn=\eta_{\mu\nu}$,
the dRGT model, on which subsequent developments are based. The
construction was based on a ``decoupling limit'' analysis (developed
for the purpose \cite{ArkaniHamed:2002sp,Creminelli:2005qk}) which
guaranteed the absence of ghost in that limit. In the perturbative
approach, it becomes difficult to extend the analysis beyond the
decoupling limit, although \cite{deRham:2010kj} outlined an argument
to show the absence of ghost in the Hamiltonian \cite{Arnowitt:1962hi}
formalism to quartic order in $h_{\mu\nu}=\gmn-\eta_{\mu\nu}$.

To proceed any further, one had to first insure that the BD ghost is
indeed absent in the dRGT model for a nonlinear $\gmn$. One also needs
to find out if the more natural case of massive gravity with a
non-flat reference metric, $\fmn\neq\eta_{\mu\nu}$, is ghost-free. It
turns out that the ``decoupling limit'', wielded powerfully in
\cite{deRham:2010ik, Creminelli:2005qk}, is not adequate to address
these situations, despite some claims to the contrary in
\cite{Mirbabayi:2011aa} and some of its citations.\footnote{The
  decoupling limit analysis does not extend beyond the decoupling
  limit for two obvious reasons: (1) It involves working with
  St\"uckelberg fields $\phi^a$, introduced via
  $\fmn=\p_\mu\phi^a\p_\nu\phi^b\eta_{ab}$, rather than with the
  metric $\gmn$. The $\phi^a$ mix only with the 4 ``gauge'' modes of
  $\gmn$ under coordinate transformations and learn about the
  potential BD ghost through them. But the BD ghost mostly resides in
  the remaining 6 components of the metric and cannot be completely
  transferred to the St\"uckelberg fields by coordinate
  transformations (otherwise, ghost fluctuations would be expressible
  as $\nabla_{(\mu}\xi_{\nu)}$ and would not contribute to
  interactions between conserved sources). Similarly, it is incorrect
  to argue that the number of independent modes in $\gmn$ can be
  reduced to 2 simply by coordinate transformations, by citing the
  analogy with GR. In GR this counting is done on-shell and holds only
  for the solutions of the massless Einstein's equations. Such a
  counting does not hold for the massive gravity equations. It has
  also been argued that one can transform $\gmn$ to $\eta_{\mu\nu}$ by
  a coordinate transformation and then apply the decoupling limit.
  This argument ignores the elementary fact that one cannot choose a
  locally flat coordinate systems over the entire spacetime.
  Obviously, the flat space action and equations of motion are not the
  same as the curved space ones. Thus, away from the decoupling limit
  it is not enough to study the $\phi^a$ alone, ignoring $\gmn$ and
  the potential ghost within it, by invoking the above arguments. (2)
  So far, it is not obvious how to obtain a decoupling limit for a
  generic non-flat $\fmn$. For example, see \cite{deRham:2012kf} for a
  recent attempt to find a decoupling limit for a de Sitter $\fmn$.}

The nonlinear analysis that can answer the above questions was
developed in \cite{Hassan:2011hr,Hassan:2011tf} based on the formalism
of \cite{Hassan:2011vm}. In \cite{Hassan:2011hr}, for the case
$\fmn=\eta_{\mu\nu}$, it was proven that the dRGT model was ghost-free
at the nonlinear level. This conclusively established the absence of
the BD ghost for the first time. The theory with generic non-flat
$\fmn$ was considered first in \cite{Hassan:2011tf} and also proven to
be ghost-free nonlinearly. This generic $\fmn$ theory provides the
most natural setup for discussing massive gravity. But from the point
of view of this paper, it describes rather a massive spin-2 field
$\gmn$ in a non-dynamical gravitational background $\fmn$.

Finally, \cite{Hassan:2011zd} obtained the ghost-free bimetric theory
for two interacting spin-2 fields $\gmn$ and $\fmn$ with the correct
kinetic structure, by exploiting the symmetries of the interactions.
An issue raised in \cite{Kluson:2011qe} about the existence of a secondary
constraint that was needed for the consistency of the formalism was
cleared up in \cite{arXiv:1111.2070}. In this paper we will work
mostly with this Hassan-Rosen bimetric theory. For related work, see
\cite{Chamseddine:2011mu,Hassan:2012qv,Kluson:2012wf,Baccetti:2012bk,
Baccetti:2012re,Khosravi:2011zi,Khosravi:2012rk,Saridakis:2012jy,
Cai:2012ag,Gumrukcuoglu:2011zh,Crisostomi:2012db, Berg:2012kn,
Paulos:2012xe}.   

The rest of the paper is organized as follows. In section 2, we review
the ghost-free bimetric theory and discuss the proportional background
solutions. In section 3, we obtain the linear mass eigenstates and
compute the general expression for the FP mass. We also discuss the
weak gravity limit. In section 4, we obtain the nonlinear massless and
massive spin-2 modes and show that the massless mode does not have
ghost-free minimal matter couplings. In section 5, we express the
bimetric action in terms of $g$ and the massive mode $M$ and discuss
some of its features. We also discuss the coupling of the massive
spin-2 field $M$ to matter. In section 6, the discussion is extended
to multimetric theories. Section 7 contains a brief discussion of the
results and some comments. Appendix A summarizes some useful equations
used in the text. Appendix B describes a rescaling that render the
action more symmetric. Finally appendix C contains the details of the
bimetric action in terms of the nonlinear massless and massive modes
$G$ and $M^G$.

\section{Proportional-background solutions in bimetric theory}
Generic solutions of the bimetric theory have little resemblance to 
solutions in general relativity. In this section we concentrate on a
particular class of bimetric background solutions that are
indistinguishable from backgrounds in general relativity. Although
very restrictive, this helps in identifying bimetric theories that are
close to general relativity. The solutions are also useful in
analyzing the linear and nonlinear mass spectrum of the bimetric
theory. We begin with a review of the ghost-free bimetric action.

\subsection{Review of the ghost-free bimetric theory} 
\label{rev}
Here we briefly review the ghost-free bimetric action and equations of
motion. The ghost-free bimetric action, excluding matter couplings,
is \cite{Hassan:2011zd},
\be
S_{gf}=\!\!\int\td^4x\left[m_g^2\sqrt{-\det g}\,R_g
+m_f^2 \sqrt{-\det f}\, R_f
- 2 m^4\sqrt{-\det g}\,
V\Big(\sqrt{g^{-1} f}\,\,;\,\beta_n\Big)\right].
\label{action} 
\ee
The potential $V$ is given by,
\be
V\left(\mathbb{X}\mathrm{;}\,\beta_n\right)
=\sum_{n=0}^4\beta_n\,e_n\left(\mathbb{X}\right)\,,
\label{int_pot}
\ee
where, $e_n(\mathbb{X})$ are elementary symmetric polynomials of the
eigenvalues of the matrix $\mathbb{X}$. In 4 dimensions they can be
expressed as,    
\be
e_{0}= 1,\hspace{.3cm}
e_{1} = [\mathbb{X}], \hspace{.3cm}  
e_{2} = \tfrac1{2}([\mathbb{X}]^2-[\mathbb{X}^2]),\hspace{.3cm}  
e_{3}=\tfrac1{6}([\mathbb{X}]^3
-3[\mathbb{X}][\mathbb{X}^2]+2[\mathbb{X}^3]),\hspace{.3cm}    
e_{4}= \det(\mathbb{X})\,,
\label{esp}
\ee
where, $[~]$ denotes the matrix trace. This potential $V$ was first
suggested, for $\fmn=\eta_{\mu\nu}$ in
\cite{deRham:2010ik,deRham:2010kj} as the unique candidate for a
ghost-free massive gravity, based on a ``decoupling limit''
analysis. That it was ghost-free nonlinearly was proven in
\cite{Hassan:2011hr}. The theories with general and dynamical $\fmn$
where first considered and shown to be ghost-free in \cite{
  Hassan:2011vm,Hassan:2011tf,Hassan:2011zd,arXiv:1111.2070}. The
square root matrix $\mathbb{X}=\sqrt{g^{-1}f}$ in $V$ is necessary to
avoid the ghost, but also complicates the analysis.

The independent parameters in the action (\ref{action}) are the five
dimensionless $\beta_n$ and the two ``Planck masses'', $m_g$ and
$m_f$. The mass scale $m$ is degenerate with the $\beta_n$ and can be
expressed in terms of the other mass parameters. Integrating out
matter fields coupled to the $g$ and $f$ metrics respectively, results
in vacuum energy contributions to $\beta_0$ and $\beta_4$. The
remaining $\beta_n$ measure the strength of nonlinear interactions
between the two metrics. An important property of $V$ is,     
\be
\sqrt{-\det g}\,\,V(\sqrt{g^{-1}f}\,\mathrm{;}\,\beta_n)  
=\sqrt{-\det f}\,\,V(\sqrt{f^{-1}g}\,\mathrm{;}\,\beta_{4-n}) \,.  
\label{A_finvgpot} 
\ee 
Then, the action (\ref{action}) is symmetric under the simultaneous   
replacements, 
\be  
g\leftrightarrow f\,,\qquad \beta_n \rightarrow\beta_{4-n}\,,\qquad
m_g\leftrightarrow m_f\,.  
\label{f-g}
\ee
The $\gmn$ and $\fmn$ equations of motion with generic
``matter'' couplings are \cite{Hassan:2011vm},    
\begin{align}
\label{g_eom}
&R_{\mu\nu}(g)-\tfrac{1}{2}\gmn R(g)+\tfrac{m^4}{m_g^{2}}\,
V_{\mu\nu}^{g}= \tfrac{1}{m_g^{2}}\,T^{g}_{\mu\nu}\,, \\
&R_{\mu\nu}(f)-\tfrac{1}{2}\fmn R(f)+\tfrac{m^4}{m_f^{2}}\,
V_{\mu\nu}^{f}= \tfrac{1}{m_f^{2}}\,T^{f}_{\mu\nu}\,.
\label{f_eom}
\end{align}
Here, the stress-energy tensors are defined by $T^{g}_{\mu\nu}=-(1/\sqrt{g})\,\delta S_m/\delta g^{\mu\nu}$, and similarly for $T^{f}_{\mu\nu}$,
where $S_m$ is the matter action added to $S_{gf}$ as in \eqref{action-gf_rev}. The interaction contributions $V_{\mu\nu}^g$ and $V_{\mu\nu}^f$ are explicitly given by, 
\begin{align}
V_{\mu\nu}^g=\sum_{n=0}^3(-1)^n\beta_n\,\,g_{\mu\lambda}\,
Y_{(n)\nu}^\lambda(\sqrt{g^{-1}f})\,,\qquad
V_{\mu\nu}^f=\sum_{n=0}^3(-1)^n\beta_{4-n}\,\,f_{\mu\lambda}\, 
Y_{(n)\nu}^\lambda(\sqrt{f^{-1}g})
\label{V}
\end{align}
where the matrices $Y_{(n)\nu}^\mu(\mathbb{X})$ can be expressed as,          
\be
Y_{(n)}(\mathbb{X})=\sum_{r=0}^n (-1)^r \, \mathbb{X}^{n-r}
\,e_r(\mathbb{X}).
\ee
The two expressions in (\ref{V}) are related through the replacements 
(\ref{f-g}). In this sense, the bimetric theory treats
$g_{\mu\nu}$ and $f_{\mu\nu}$ on the same footing.
%

The usual Bianchi identities of the curvature tensors together with
$\nabla_{g,f}^\mu T_{\mu\nu}^{g,f}=0$, imply the Bianchi constraints,
which are independent of the scales $m_g$ and $m_f$,  
\be
\nabla^\mu_{g} V_{\mu\nu}^{g}=0\,,\qquad
\nabla^\mu_{f} V_{\mu\nu}^{f}=0\,.
\label{Bianchi}
\ee


\subsection{Proportional background solutions}\label{propback}

Generic solutions of the bimetric theory are very different from
solutions in general relativity. Here we consider a particular class
of bimetric solutions $\bgmn$ and $\bfmn$, sourced by $\bar
T^g_{\mu\nu}$ and $\bar T^f_{\mu\nu}$, which coincide with solutions
for the metric in GR. These are solutions of the type\footnote{For
  non-proportional metrics, GR type solutions exist for certain
  choices of $\beta_n$, but for specific metric ansatz, say the FRW
  ansatz \cite{vonStrauss:2011mq, Comelli:2011wq}.}
\be 
\bfmn=c^2\bgmn \,, \label{prop-bg}
\ee
and exist only if $\bar T_{\mu\nu}^f\propto \bar T_{\mu\nu}^g$. This
restriction on the matter sources is not always realistic, but such 
solutions are motivated by other considerations discussed at the end
of this section.   

For the ansatz (\ref{prop-bg}), the Bianchi constraints
(\ref{Bianchi}) imply that $c$ is a constant. Then (\ref{g_eom}) and
(\ref{f_eom}) reduce to two copies of Einstein's equations for the
curvatures of $\bgmn$,      
\be
\bar R_{\mu\nu}-\tfrac1{2}\bgmn\bar R +\Lambda_g\,\bgmn =
\tfrac{1}{m_g^2}\bar T^{g}_{\mu\nu} \,, \qquad
\bar R_{\mu\nu}-\tfrac1{2}\bgmn \bar R+\Lambda_f\,\bgmn =
\tfrac{1}{m_f^2}\bar T^{f}_{\mu\nu}\,, 
\label{bgf_eom}
\ee
where the cosmological constants are given by,
\be
\Lambda_g=\tfrac{m^4}{m_g^2}\left(\beta_0+3c\beta_1+3c^2\beta_2+
c^3\beta_3\right)\,, \quad    
\Lambda_f = \tfrac{m^4}{m_f^2 c^2} \,\left(c\beta_1 +3c^2\beta_2 +
3c^3\beta_3+c^4\beta_4\right)\,.
\label{Lambda_gf}
\ee
Obviously, the equations are consistent only if,
\be
\left(\Lambda_g-\Lambda_f\right)\bgmn = 
\left(m_g^{-2}\,\bar T^{g}_{\mu\nu}-m_f^{-2}\,\bar T^{f}_{\mu\nu}
\right)\,. 
\label{bgrel}
\ee
The vacuum energy contributions to $\bar T^g_{\mu\nu}$ and
$\bar T^f_{\mu\nu}$ can always be absorbed in $\beta_0$ and $\beta_4$.
Hence, the right-hand side can be assumed to contain no piece
proportional to $\bgmn$. Then, for localizable sources, each side of the
above  equation must vanish separately,\footnote{If
  $\Lambda_g\neq\Lambda_f$, then (\ref{bgf_eom}) and (\ref{bgrel}) 
  would lead to a complicated differential equation for the
  $T_{\mu\nu}$'s.}
\be
\Lambda_g=\Lambda_f\,,\qquad 
\bar T^f_{\mu\nu}=\tfrac{m_f^{2}}{m_g^{2}}\,\bar T^g_{\mu\nu}\,.
\label{constcosconst}
\ee
Later it will be seen that (\ref{constcosconst}) is also crucial for
the existence of spin-2 massive and massless eigenstates.

The equation $\Lambda_g=\Lambda_f$ determines the constant $c$ in
terms of the parameters of the theory, $m_g$, $m_f$ and the five
$\beta_n$, through the quartic equation,
\be
\alpha^2 \beta_3 c^4+ (3\alpha^2 \beta_2-\beta_4)c^3   
+3(\alpha^2 \beta_1 -\beta_3)c^2 +(\alpha^2 \beta_0-3\beta_2)c
-\beta_1 =0 
\label{c-gen}
\ee 
where
\be
\alpha=\frac{m_f}{m_g}\,.
\label{alpha}
\ee
On solving the above equation for $c$ and substituting in $\Lambda_g$,
one obtains the cosmological constant in terms of the parameters of
the theory. 

To get a feeling for the behaviour of $c$ we can solve this equation
for the simple case of the ``minimal'' bimetric model, corresponding
to $\beta_1=\beta_3=0$. Then (\ref{c-gen}) gives,
\be
c^2=\frac{3\beta_2-\alpha^2\beta_0}{3\alpha^2\beta_2 -\beta_4}
\,,\qquad (\beta_1=\beta_3=0)\,,
\label{c-min}
\ee
showing that, in general, $c^2$ could have any value depending on
$\beta_0, \beta_2$ and $\beta_4$ . This in turn gives the cosmological  
constant, 
\be
\Lambda_g=\Lambda_f=\frac{m^4}{m_g^2} \,\,
\frac{9\beta_2^2-\beta_0\beta_4}{3 \alpha^2 \beta_2
  -\beta_4}\,,\qquad  (\beta_1=\beta_3=0)\,.
\label{CC-min}
\ee
The scale $m^4$ can be eliminated in terms of the Fierz-Pauli mass of
the massive excitation given in the next subsection.

Note that the most general set of parameters for which the theory 
admits flat space as a solution is obtained from $\Lambda_g=0$, after
solving for $c$. This condition can be solved for one of the $\beta_n$,
leaving the rest free. In contrast, specifying flat space through
$\bar f=\bar g=\eta$ and $\Lambda_g=\Lambda_f=0$ will eliminate two of
the $\beta_n$ leading to a smaller parameter space. For example, in
the minimal case considered above, theories that admit flat space as a
background are parameterized by $9\beta_3^2=\beta_0\beta_4$, whereas
forcing $c=1$ gives the smaller parameter space
$\beta_0=\beta_4=-3\beta_3$.     

\subsection{Discussion}

If one interprets one of the two metrics, say $\bar g_{\mu\nu}$, as
the gravitational metric coupled to ordinary matter $\bar T^g$ with
Planck mass $m_g$, then one recovers all classical backgrounds of
GR. However, the requirement $\bar T^f_{\mu\nu} = \alpha^2\bar
T^g_{\mu\nu}$ imposed by these solutions is not realistic 
(except possibly for $\alpha=1$ such that the two metrics are coupled to the
same matter). In spite of this such backgrounds are motivated by other 
considerations.

(1) The ansatz (\ref{prop-bg}) results in the most general class of
bimetric backgrounds for which there exist a well-defined massive
spin-2 fluctuation $\delta M_{\mu\nu}$ with a Fierz-Pauli structure,  
\be
\label{FPstruc} 
\mfp\sqrt{-\det\bar g}\left[\delta M^\mu_{\,\,\nu}\delta
M^\nu_{\,\,\mu}-(\delta M^\mu_{\,\,\mu})^2\right]\,, 
\ee
along with a decoupled massless spin-2 fluctuation $\delta
G_{\mu\nu}$, as will be discussed below. The explicit expressions help
extend the linear mass eigenstates $\delta M$ and $\delta G$ to
nonlinear fields $M$ and $G$. The $c\neq 1$ case helps in 
identifying $G$. Finally, we interpret the bimetric theory as a   
nonlinear theory of a massive spin-2 field coupled to gravity. 

(2) To be consistent with observations, the solutions for the
nonlinear field that is identified with the gravitational metric must
be very close to the corresponding solutions in GR. More specifically,
corrections to the GR solutions coming from the non-gravitational
sector must be strongly suppressed in the weak gravity limit. These
issues are difficult to investigate nonlinearly. Thus as a first
step, one can consider perturbations around backgrounds of the type
(\ref{prop-bg}), sourced by independent $\delta T^g$ and $\delta T^f$
to probe parameter regions that suppress deviations from GR. Later we
will also see that the nonlinear massive field defined with respect to
these backgrounds probes deviations from GR.

\section{Linear massive and massless modes }

In Minkowski backgrounds $\bar g=\bar f=\eta$, where the concept of
mass is well defined through the Poincar\'e group, the spectrum of
bimetric theory is known to consist of a massive and a massless spin-2
fluctuation \cite{ISS,Hassan:2011zd}. Such backgrounds exist
only after two out of the five $\beta_n$ parameters are fixed. Here we
consider the spectrum of linear fluctuations in the theory with
arbitrary $\beta_n$. In non-flat backgrounds we define a massive fluctuation
as one with a Fierz-Pauli mass term (\ref{FPstruc}). In
bimetric theory, such mass terms arise only around proportional
backgrounds $\bfmn=c^2\bgmn$ considered above. For independent source
fluctuations $\delta T^g_{\mu\nu}$ and $\delta T^f_{\mu\nu}$ the
expressions help in characterizing deviations from GR. The linear mass 
eigenstates are extended to nonlinear fields in the next section. 

\subsection{Massive and massless modes in the linearized theory} 
\label{app_lin_bimetric}

Consider canonically normalized fluctuations around the
$\bfmn=c^2\bgmn$ backgrounds,    
\be
g_{\mu\nu}=\bar g_{\mu\nu}+\tfrac{1}{m_g}\,\delta g_{\mu\nu}\,,\qquad
f_{\mu\nu}=c^2\bar g_{\mu\nu}+\tfrac{c}{m_f}\, \delta f_{\mu\nu}\,.\qquad
\label{pert}
\ee
Then to linear order, 
\be
(\sqrt{g^{-1}f})^\rho_{\ph\rho\nu}=c\,\delta^\rho_\nu
+\dS^\rho_{\ph\rho\nu}\,, \quad {\rm where},
\quad \dS^\rho_{\ph\rho\nu} = \frac{1}{2m_f}\,
\bg^{\rho\mu}\left(\df_{\mu\nu}-c\,\tfrac{m_f}{m_g}\,\dg_{\mu\nu}\right)\,.
\label{deltaM}
\ee
Expanding the interaction contributions (\ref{V}) and using the
results in appendix 
(\ref{app_Rrels}) gives the linearized equations,
\begin{align}
\bar{\mathcal{E}}_{\mu\nu}^{\rho\sigma}\,\delta g_{\rho\sigma}
+\Lambda_g\delta\gmn -\tfrac{m^4 B}{m_g}\,
\bg_{\mu\rho}\left({\dS^\rho}_\nu-\delta^\rho_\nu{\dS^\sigma}_{\sigma}
\right)
&= \tfrac1{m_g}\delta T^{g}_{\mu\nu}\,,
\label{geomlin} \\
\bar{\mathcal{E}}_{\mu\nu}^{\rho\sigma}\,\delta f_{\rho\sigma}
+\Lambda_f \delta f_{\mu\nu}+ \tfrac{m^4 B}{c m_f}\,
\bg_{\mu\rho}\left({\dS^\rho}_\nu-\delta^\rho_\nu{\dS^\sigma}_{\sigma}
\right)
&= \tfrac{1}{m_f}\delta T^{f}_{\mu\nu}\,,
\label{feomlin}
\end{align}
where,
\be
B=\tfrac{1}{c}(c\beta_1+2 c^2\beta_2 + c^3\beta_3)\,.
\ee
$\bar{\mathcal{E}}$ is given in (\ref{A_ERdef}). 
By taking appropriate linear combinations, (\ref{geomlin}) and
(\ref{feomlin}) can be easily decoupled in terms of a massive 
($\dMg_{\mu\nu}\sim\bar g_{\mu\lambda}\dS^\lambda_{\,\,\nu}$) and a
massless $\delta G_{\mu\nu}\sim\delta g_{\mu\nu}+c\,(m_f/m_g)
\delta f_{\mu\nu}$ spin-2 fluctuation. However, this is possible only
if $\Lambda_g=\Lambda_f$, which was also required on other
grounds. Finally, the canonically normalized massless and massive 
fluctuations become,\footnote{The canonical normalization is
  determined from the action requiring that  
$\delta g\bar{\mathcal{E}}\delta g+\delta f\bar{\mathcal{E}}\delta f=
\dMg\bar{\mathcal{E}}\dMg+\delta G\bar{\mathcal{E}}\delta G$. This
value will change if $\bar{\mathcal{E}}$ on the rhs is computed with
the background metric $\bar G$ instead of $\bg$.}
\begin{align}
\delta G_{\mu\nu}&= \frac{1}{\sqrt{c^2\alpha^2 +1}}
\left(\delta g_{\mu\nu}+ c\alpha\,\delta f_{\mu\nu}\right)\,, 
\label{dGc} \\
\dMg_{\mu\nu}&=\frac{1}{\sqrt{c^2\alpha^2 +1}}
\left(\df_{\mu\nu}- c\alpha \,\dg_{\mu\nu} \right)\,,
\label{dMc}
\end{align}
where $\alpha=\frac{m_f}{m_g}$. The
corresponding massless and massive equations are,    
\begin{align}
&\bar{\mathcal{E}}_{\mu\nu}^{\rho\sigma}\,\delta G_{\rho\sigma}
+\Lambda_g \delta G_{\mu\nu} = 
\frac{\delta T^{(g)}_{\mu\nu}+c^2\,\delta
  T^{(f)}_{\mu\nu}}{m_g\sqrt{c^2\,\alpha^2+1}}\,, 
\label{Glineq}  \\
& \bar{\mathcal{E}}_{\mu\nu}^{\rho\sigma}\,\dMg_{\rho\sigma}
+\Lambda_g \dMg_{\mu\nu} +\tfrac{m_{\mathrm{FP}}^2}{2}\,
\left(\dMg_{\mu\nu}-\bg_{\mu\nu}\bg^{\rho\sigma}
\dMg_{\rho\sigma}\right)
=c\,\frac{\delta T^{(f)}_{\mu\nu}-\alpha^2 \delta T^{(g)}_{\mu\nu}}
{m_f\sqrt{c^2\alpha^2+1}}\,. 
\label{Mlineq}
\end{align}
The Fierz-Pauli mass above is parameterized as,
\be
m^2_{\mathrm{FP}}=m^4 (c\beta_1+2 c^2\beta_2 + c^3\beta_3)
\Big(\frac{1}{c^2 m_f^2} + \frac{1}{m_g^2}\Big)\,.
\ee

From (\ref{Glineq}) it is evident that, in the
background metric $\bgmn$, the massless fluctuation $\delta
G_{\mu\nu}$ couples to matter with the effective Planck mass,  
\be
m_p=m_g \sqrt{c^2\alpha^2+1}=\sqrt{m_g^2+c^2 m_f^2}\,.
\ee 
which must be large for gravity to be weak. This can be achieved in
different ways with different consequences. It is also evident that at
the linear level, $\delta G$ behaves like the metric perturbation in
GR. Deviations from GR emerge mainly at the nonlinear level.

Away from proportional backgrounds, the fluctuations generically do
not have a Fierz-Pauli mass term. The analysis is further complicated 
by the fact that in such cases, $\sqrt{g^{-1}f}$ does not have a
simple expansion. 

\subsection{Weak gravity limit} \label{wglsec}

In GR, gravity is described in terms of a massless spin-2 field
minimally coupled to matter, as required by the weak equivalence
principle. Considering the observational evidence in support of GR, it
is natural to assume that in interacting spin-2 theories too, the
gravitational interactions must be associated predominantly, if not
exclusively, with the massless spin-2 mode of the theory. The validity
of the weak equivalence principle then requires that this
gravitational mode must couple to matter in more or less the same way
that the gravitational metric couples to matter in GR. This simple
observation leads to the following possibilities.

(1) Let's assume that the massless mode $\delta G_{\mu\nu}$ can be
extended to a nonlinear field $G_{\mu\nu}$. If $G_{\mu\nu}$
could directly couple to matter in a ghost-free manner, using the same
minimal coupling prescription as in GR, then such matter couplings
would not directly violate the weak equivalence principle. In this
case, one should express the operator
$\bar{\mathcal{E}}_{\mu\nu}^{\rho\sigma}$ in (\ref{Glineq}) 
in terms of the background $\bar G_{\mu\nu}$ which will be
proportional to $\bgmn$, giving, 
$\bar{\mathcal{E}}_{\mu\nu}^{\rho\sigma}(\bar g)=a(c,m_f,m_g)\,
\bar{\mathcal{E}}_{\mu\nu}^{\rho\sigma}(\bar G)$. Then the Planck mass
is $a m_p$ which must be large. Later we identify a nonlinear
massless mode $G_{\mu\nu}$ and show that it cannot couple to matter in
a ghost-free way. In the absence of 
consistent direct couplings of $G_{\mu\nu}$ to matter, a different
approach is needed.  

(2) Now consider setups where matter fields can directly couple only
to the metrics $\gmn$ or $\fmn$ (as in (\ref{action-gf-m_rev})), but not to
$G_{\mu\nu}$. This would be a natural way of accommodating the weak
equivalence principle only if $\gmn$ or $\fmn$ described gravity. On
the other hand, empirically, gravity is well described by a massless
spin-2 field, which in the bimetric setup is $G_{\mu\nu}$. These two
requirements can be reconciled if the massless mode $G_{\mu\nu}$ is
dominated by $\gmn$ or $\fmn$. Here we consider the possibility that
$G_{\mu\nu}$ is mostly made up of $\gmn$.\footnote{Equally well, one
  could replace $\gmn$ by $\fmn$ and $c$ by $1/c$} The limits in which
this holds can be identified at the linearized level from (\ref{dGc}),
where, $\delta G_{\mu\nu}\sim \delta\gmn$ holds in the limit     
\be
m_g >> c\, m_f \,.
\label{mgcmf}
\ee
This can be achieved by a small $m_f$ or a small $c$ of both. Whether
this choice is natural or not, will not be addressed here. Also in
this limit, the massive fluctuation $\delta M_{\mu\nu}$ is mostly
saturated by $\fmn$.  The strength of $\delta M_{\mu\nu}$ interactions
depend on the relative values of $c$ and $m_f$. Following this
reasoning, in section \ref{sec_nonlin_M} we consider the nonlinear
action in terms of $\gmn$ and the nonlinear massive mode $M_{\mu\nu}$. 

\section{The nonlinear massless and massive modes}
Now we consider extending the mass eigenstates of linearized bimetric
theory to nonlinear fields. The ADM analysis of the bimetric action
shows that even nonlinearly the theory has seven propagating modes
\cite{Hassan:2011zd}. But only their linear fluctuations around
$\bfmn=c^2\bgmn$ backgrounds combine into well defined massless and
massive spin-2 states. Here we explore the nonlinear extensions of
these mass eigenstates. In a theory with general covariance, spin-2
fields are minimally represented by rank-2 symmetric tensors. Below we
find such tensors that reduce to the mass eigenstates (\ref{dGc}) and
(\ref{dMc}) at the linear level. Since this choice is not unique, one
can also invoke simplicity as a criterion. These are the only criteria
employed here. We have not considered if the nonlinear modes also
propagate two, respectively, five degrees of freedom at the nonlinear
level. 

\subsection{The nonlinear massless spin-2 field
  \texorpdfstring{$G$}{G}} 
\label{sec_masslessansatz}

The nonlinear massless mode is a symmetric rank-2 tensor
$\Gmn$ that reproduces the massless fluctuation $\delta G$ (\ref{dGc})
at the linear level. To determine $\Gmn$ we work with the $(1,1)$
tensor, 
\be
S^\mu_{\,\,\nu}= \big(\sqrt{g^{-1}f}\,\big)^\mu_{\,\,\,\,\nu}\,.
\ee
First note that $\sqrt{g^{-1}f}=g^{-1}(\sqrt{fg^{-1}})\,g$. This
follows on writing $\sqrt{g^{-1}f}=\sqrt{{\mathbb 1}+(g^{-1}f-
{\mathbb 1})}$ and formally expanding the square-root. We then have
the important property,\footnote{From the above properties of
  $S^\mu_{~\nu}$ it follows that, $f=S^{\mathrm{T}}gS$. 
Hence, $S$ is a local transformation between $\fmn$ and $\gmn$, or a 
generalized vielbein. Further, in terms of $S$ the proportional
backgrounds (\ref{prop-bg}) are characterized by the background value 
$\bar S^\mu_{\,\,\nu}= c \,\delta^\mu_{\,\,\,\nu}$
which is invariant under general coordinate transformations.} 
\be
gS =S^T g\,.
\label{gM-sym}
\ee
Now, let us start with a general symmetric $(0,2)$ tensor $\Gmn(g,f)$
and, in it, replace $f$ by $S$ through $f=gS^2$. In general, $\Gmn$
could contain powers of $S^\mu_{\,\,\nu}$, $(S^{-1})^\mu_{\,\,\nu}$,
$(S^T)^{\,\,\mu}_{\nu}$ and $(S^{-1T})^{\,\,\mu}_{\nu}$ contracted
with $g_{\mu\nu}$ in the right way to produce a $(0,2)$ tensor. Then
using (\ref{gM-sym}) it is easy to see that general covariance alone
restricts $\Gmn$ to the form, 
\be
G = g\, \Phi(S)\,,
\label{GPhi}
\ee
where $\Phi^\mu_{\,\,\nu}$ is a matrix function of the matrix
$S^\mu_{\,\,\nu}$ and its inverse, but not its transpose. On the
proportional backgrounds $\bar f=c^2\bg$, where $\bar S^\mu_{\,\,\nu}
=c\,\delta^\mu_{\,\,\nu}$, this becomes,    
\be
\bar G = \bar g\,\bar\Phi\equiv\bar g\,\Phi({\bar S=c\,\mathbb 1})\,.
\ee 
Clearly, $\bar\Phi=\phi(c){\mathbb 1}$ for a scalar
$\phi(c)$. $\bar\Phi(\bar S)$ depends on $c$ in two ways: through an
explicit dependence of $\Phi$ on $c$ (e.g., through normalizations), 
and through $\bar S$. If these two types of contributions could be
disentangled, $\Phi$ could be uniquely reconstructed from $\bar\Phi$.    

Let us now consider fluctuations $\Gmn=\bar G_{\mu\nu}+\delta
\Gmn'$. These can be computed using the canonically normalized
variables of the previous section. But to ensure explicitly that the 
equations depend on $c$ only through $\bar S$ and not through
normalizations, here we work with, 
\be
\gmn=\bgmn + \delta\gmn'\,,\qquad \fmn=\bfmn + \delta\fmn'\,,\qquad
\delta S^\mu_{\,\,\nu}=\tfrac{1}{2}\bar g^{\mu\lambda}(\tfrac{1}{c}\delta
f'_{\lambda\nu} - c\, \delta g'_{\lambda\nu})\,,
\label{pert-nc}
\ee   
Then the fluctuation of the nonlinear massless field becomes, 
\begin{align}
\delta\Gmn' &=\delta g'_{\mu\lambda}\bar\Phi^\lambda_{\,\,\nu}+
\bar g_{\mu\lambda}\frac{\p\Phi^\lambda_{\,\,\nu}}{\p
  S^\alpha_{\,\,\beta}}\Big\vert_{\bar S} \delta S^\alpha_{\,\,\beta}
+\cdots  \nn\\
&=\delta g'_{\mu\lambda}\bar\Phi^\lambda_{\,\,\nu}
-\frac{c}{2}\bar g_{\mu\lambda}\frac{\p\Phi^\lambda_{\,\,\nu}}{\p
S^\alpha_{\,\,\beta}}\Big\vert_{\bar S}\bar g^{\alpha\sigma}
\delta g'_{\sigma\beta}
+\frac{1}{2c}\bar g_{\mu\lambda}\frac{\p\Phi^\lambda_{\,\,\nu}}{\p
  S^\alpha_{\,\,\beta}}\Big\vert_{\bar S}\bar g^{\alpha\sigma}
\delta f'_{\sigma\beta} + \cdots
\label{dGnl}
\end{align}
On the other hand, in terms of (\ref{pert-nc}) the massless
fluctuation (\ref{dGc}) becomes,  
\be
\delta\Gmn'= A \left(\delta\gmn'+\alpha^2 \delta\fmn'\right)\,,
\label{dGnc}
\ee
with a normalization $A$. Comparing the coefficients of the
fluctuations in (\ref{dGnl}) and (\ref{dGnc}) gives two equations 
for $\Phi$ as a function of $S$, evaluated at $\bar S= c{\mathbb 1}$, 
\begin{align}
A^{-1}\,\bar\Phi^\lambda_{\,\,\nu}= (1+\alpha^2\,c^2)\,
\delta^\lambda_{\,\,\nu}\,,\qquad 
A^{-1}\,\frac{\p\Phi^\lambda_{\,\,\nu}}{\p S^\alpha_{\,\,\beta}}
\Big\vert_{c \mathbb 1} = 2c\,\alpha^2 \delta^\lambda_{\,\,\alpha} 
\delta^\beta_{\,\,\nu}\,.
\label{Phi-fluc}
\end{align}
The right-hand sides acquire their $c$-dependence only through $\bar
S=c\mathbb 1$, and not normalizations. It is then natural to assume
that $A^{-1}\Phi(S)$ depends on $c$ only through 
$\bar S=c\mathbb 1$. This leads to the unique solution,
\be
\Phi= A\,({\mathbb 1}+\alpha^2\,S^2)\,,
\label{Phi}\\
\ee
obtained from the first equation on replacing $c$ by $S$. It gives   
the nonlinear massless mode,
\be
\Gmn = A\,(\gmn + \alpha^2\,\fmn)\,.
\label{Gnl}
\ee
Without loss of generality we can set $A=1$. The fluctuations of this
mode can be canonically normalized either with respect to $\bar g$, to
give (\ref{dGc}), or with respect to $\bar G_{\mu\nu}$. 

Other nonlinear extensions of $\delta G_{\mu\nu}$ can be found if
$A^{-1}\Phi(S)$ is allowed to have an explicit dependence on $c$,
besides that coming from $\bar S$. To find these note that,   
\be
\frac{\delta (S^n)^\lambda_{\,\,\nu}}{\delta S^\alpha_{\,\,\beta}}
\Big\vert_{c\mathbb 1}=n\, c^{n-1}\,\delta^\lambda_{\,\,\alpha} 
\delta^\beta_{\,\,\nu}\,.
\label{dMn}
\ee
Then general c-dependent solutions of (\ref{Phi-fluc}) can be written   
as,    
\be
\Phi= A\Big(a_0 + 2\alpha^2 \sum_{n>0} a_n\,c^{2-n}\,S^n
\Big)\,,\quad \mathrm{with}\quad
a_0+2\alpha^2c^2\sum_{n>0} a_n =1+\alpha^2c^2\,,\quad
\sum_{n>0} n\,a_n=1\,. 
\ee
Of course, an infinite number of such solutions exists, the one with the
lowest power of $S$ being $A^{-1}\Phi=1-\alpha^2c^2+2\alpha^2cS$. A
non-polynomial solution of (\ref{Phi-fluc}) is $A^{-1}\Phi=(1+
\alpha^2c^4S^{-2})^{-1}$, giving the massless mode $G^{-1}=g^{-1}+
\alpha^2c^4 f^{-1}$.

Of all these, the c-independent solution (\ref{Gnl}) gives the
simplest invertible relation between the nonlinear modes and the
original bimetric variables $g$ and $f$. Note that to identify this
unique c-independent mode, it was important to work with $c\neq
1$ backgrounds. Otherwise, at $c=1$, this criterion is not useful.

\subsection{The nonlinear massive spin-2 field
  \texorpdfstring{$M$}{M}}  
\label{sec_massiveansatz}

From the outset it is evident that the nonlinear massive field is
closely related to $S^\mu_{\,\,\nu}= \big(\sqrt{g^{-1}f}
\big)^\mu_{\,\,\,\,\nu}$. This is hinted by the linear equations
(\ref{deltaM}), (\ref{dMc}) and also by the fact that the mass
potential $V$ in (\ref{int_pot}) is a polynomial in $S$. $S^\mu_{~\nu}$
is a $(1,1)$ tensor but it can be brought to a symmetric $(0,2)$ form
in more than one way and the nonlinear extensions of the 
massive spin-2 fluctuation are related to these $(0,2)$ forms. Here we 
consider two nonlinear extensions, $M_{\mu\nu}$ and $M^G_{\mu\nu}$,
before discussing the general case.  

In terms of $S_{\mu\nu}\equiv g_{\mu\lambda}S^\lambda_{~\nu}$,
equation (\ref{gM-sym}) is the symmetry condition,
\be
S_{\mu\nu}=S_{\nu\mu}\,.
\ee
The fluctuation $\delta (S_{\mu\nu})=c\delta g_{\mu\nu} +\tilde
c\dMg_{\mu\nu}$ is a mixture of $\delta g$ and the massive mode
$\dMg$ (\ref{dMc}). Now it is obvious that a nonlinear  
massive mode can be written as, 
\be
\Mmn= B \big(g_{\mu\lambda}S^\lambda_{\,\,\,\nu}-c\gmn\big)\,,
\label{Mnl}
\ee
allowing for a normalization $B$. On proportional backgrounds,
$\bar{M}_{\mu\nu}= 0$. This is a natural vacuum value for a
non-gravitational spin-2 field in the sense that it does not break
general covariance in its vacuum. Fluctuations around this background
are the massive modes $\dMg$ (\ref{dMc}) with a Fierz-Pauli mass
term.\footnote{A note on notation: for $c=1$ and $\fmn=\eta_{\mu\nu}$,  
  $g^{-1}M$ coincides with the $K=\sqrt{g^{-1}\eta}-{\mathbb{1}}$ in
  terms of which the dRGT model is written. $K^\mu_{\,\,\nu}$ was
  engineered to produce the massive mode around flat space for $c=1$,
  while $M$ represents the massive mode around any background for
  which a Fierz-Pauli mass can be written.} The condition $gS=S^Tg$
implies that $S^\mu_{~\nu}$ and $g_{\mu\nu}$ are not independent
fields, whereas $M_{\mu\nu}$ and $g_{\mu\nu}$ can be regarded as
independent. 

A different nonlinear extension of the massive fluctuation is
obtained by using $\Gmn$ (\ref{Gnl}) instead of $\gmn$, 
\be
M^G_{\mu\nu}= \tfrac{B}{A(1+\alpha^2c^2)}
\big(G_{\mu\lambda}S^\lambda_{\,\,\nu}-c\,\Gmn\big)\,.  
\label{MGnl}
\ee
The normalization is fixed such that $\delta M^G=\delta M$. It is
easier to invert the relations and express $(g,f)$ in terms of
$(G,M^G)$ rather than in terms of $(G,M)$. For more on this see
Appendix \ref{app_nonlinaction}. 
 
In general, the massive fluctuation $\dMg$ has many possible nonlinear
extensions. More nonlinear extensions can be obtained by following a
procedure similar to the massless case. By general covariance alone,
any matrix function $M$ of $g$ and $f$ can be written as    
\be
M = g\, \Psi(S)\,. 
\ee
The fluctuations of this field,
\be
\delta\Mmn=\delta g'_{\mu\lambda}\bar\Psi^\lambda_{\,\,\nu}+
\bar g_{\mu\lambda}\frac{\p\Psi^\lambda_{\,\,\nu}}{\p
  S^\alpha_{\,\,\beta}}\Big\vert_{\bar S} \delta S^\alpha_{\,\,\beta}
+\cdots  
\label{dMnl}
\ee
should be equated to the massive fluctuation (\ref{dMc}) with
arbitrary normalization $B$, 
\be
\delta\Mmn= B\,\bar g_{\mu\lambda}\,\delta S^\lambda_{\,\,\nu}\,.
\ee
This gives, 
\be
B^{-1}\bar\Psi=\Psi(c{\mathbb 1}) =0\,,\qquad
B^{-1}\frac{\p\Psi^\lambda_{\,\,\nu}}{\p S^\alpha_{\,\,\beta}}
\Big\vert_{\bar S} =\delta^\lambda_{\,\,\alpha}\delta^\beta_{\,\,\nu} \,.
\label{Psi-bg-fluc}
\ee 
Again the $c$-dependence of the right-hand sides comes only from
$\bar S$ and not from normalizations. However, now we cannot assume
that $B^{-1}\Psi$ depends on $S$ and not explicitly on $c$ since
then $\bar\Psi(c)=0$ would imply $\Psi(S)=0$, identically. At a least
a minimal $c$-dependence is needed to get a nonlinear massive mode
with a vanishing background value. The solution with the simplest and
most natural $c$ dependence is the one corresponding to (\ref{Mnl}),
\be
\Psi^\lambda_{\,\,\nu}=B\,\left(S^{\lambda}_{\,\,\nu}
-c\,\delta^{\lambda}_{\,\,\nu}\right)\,.
\ee
More general solutions of equations (\ref{Psi-bg-fluc}), involving
higher powers of $S$, are given by,
\be
\Psi=B\sum_{n\geq 0} b_n\,c^{1-n}\,S^n\,,\qquad \mathrm{with}\qquad
\sum_{n\geq 0} b_n\,=0\,,\qquad \sum_{n\geq 1} n\,b_n=1\,.
\label{Mxgen}
\ee
For example, one can check that the massive mode $M^G$ (\ref{MGnl})
is a solution to the above equations by reading off the $b_n$ from 
$B^{-1}M^G=(1+c^2\alpha^2)^{-1}\,g\,(-c+S-c\alpha^2 S^2+\alpha^2S^3)$.

\subsection{Absence of ghost-free matter coupling of the 
  massless mode \texorpdfstring{$G$}{G}} 
\label{G-ghost} 

In the previous subsection we obtained a nonlinear generalization
$\Gmn$ (\ref{Gnl}) of the massless fluctuation of bimetric theory. It
is natural to ask if $\Gmn$ could be consistently coupled to matter in
the standard way, and be interpreted as the gravitational metric. As
in GR, such matter couplings must be consistent with the weak
equivalence principle. Here we show that minimal couplings of $\Gmn$ are
not ghost-free. The alternative, then, is to regard
$\gmn$ as the gravitational metric and rely on the weak gravity limit
discussed earlier.

Consider standard minimal couplings of $\Gmn$ to matter, for example,
to a scalar field $\phi$,
\be
{\mathcal L}_{(m,G)}=- \sqrt{-G}~G^{\mu\nu}\partial_\mu\phi\,
\partial_\nu\phi\,. 
\label{G-phi}
\ee
To see if such ghost-free couplings exist in the bimetric theory, one
can perform a Hamiltonian (ADM) analysis \cite{Arnowitt:1962hi}. We
introduce the following notation for the $3+1$ decomposition of $G$,
\beqn
G_{\mu\nu}=\begin{pmatrix}
-K^2+K^lK_l~~~~ &K_j\\
K_i& \,^3G_{ij}
\end{pmatrix}\,,
\eeqn
where $K_i=^3\!G_{ij}K^j$. Standard matter couplings of $\Gmn$ such
as (\ref{G-phi}), when written in the Hamiltonian form using  
canonically conjugate variables, are linear in $K$ and $K_i$, 
\be
{\mathcal L}_{(m,G)}=\tilde{\mathcal{L}}+K\tilde{\mathcal{C}}
+K_i\tilde{\mathcal{C}}^i\,.
\label{hamg}
\ee
If the dynamics of $\Gmn$ were described by the Einstein-Hilbert
action, $\sqrt{-G}R_G\sim \Pi^{ij}\,\p_t\,^3\!G_{ij}+K {\mathcal C} +
K_i {\mathcal C}^i$, then $K$ and $K_i$ would be Lagrange multipliers
in the full theory. Their equations of motion would result in four 
constraints that, along with gauge symmetries, would eliminate
the ghost and leave two propagating modes for $\Gmn$. However, in
bimetric theory, the nonlinear action expressed in terms of $\Gmn$ and
$M$ (or $M^G$) is complicated and it is not convenient to
carry out the ghost analysis in terms of the ADM variables of $\Gmn$.
Instead, since the bimetric analysis is already known in terms of the
ADM variables for $g$ and $f$ \cite{Hassan:2011zd}, the strategy  
here is to analyze the matter coupling (\ref{hamg}) in terms of these
variables.

To this end, we introduce the $3+1$ decompositions of $g$ and $f$,
\beqn
g_{\mu\nu}=\begin{pmatrix}
-N^2+N^lN_l~~~~ &N_j\\
N_i& ^3g_{ij}
\end{pmatrix}\,,\hspace{20pt}
f_{\mu\nu}=\begin{pmatrix}
-L^2+L^lL_l~~~~ &L_j\\
L_i& ^3f_{ij}
\end{pmatrix}\,.
\eeqn
The indices on $N_i$ and $L_i$ are raised using the inverses of
$^3g_{ij}$ and $^3f_{ij}$, respectively. It is known that in terms of
new variables $n^i$ that parameterize $N^i-L^i$ through (for details
and the form of the matrix $D$, see
\cite{Hassan:2011tf,Hassan:2011zd}),   
\be
\label{redefn}
N^i=L^i+Ln^i+N{D^i}_kn^k\,,
\ee
the bimetric theory (\ref{action}) (with no matter couplings) takes
the form \cite{Hassan:2011zd},   
\be\label{lagbim}
{\mathcal L}=\meff^2\left[\pi^{ij}\partial_t g_{ij}+p^{ij}
\partial_t f_{ij}+N\mathcal{C}_g+L\mathcal{C}_f+L_iR^i_{gf}\,\right].
\ee 
For convenience we use the scaled fields of appendix B, effectively
setting $m_g=m_f$. $\pi_{ij}$ and $p_{ij}$ are the momenta conjugate
to $g_{ij}$ and $f_{ij}$. The functions $\mathcal{C}_g$,
$\mathcal{C}_f$ and $R^i_{gf}$ are independent of $N$, $L$ and $L_i$,
but depend on $n^i$ and the remaining variables. The action has the
additional property that the $n^i$ equations of motion are independent
of $N$, $L$ and $L_i$, and determine $n^i$ in terms of the 
remaining variables. Thus, $N$, $L$ and $L_i$ are five Lagrange
multipliers whose equations, in particular, $\mathcal{C}_g=0$ and
$\mathcal{C}_f=0$, provide the constraints that render the theory
ghost-free (along with the associated secondary constraints and gauge
conditions)\cite{Hassan:2011zd,arXiv:1111.2070}. 

To emphasize, this argument for the absence of ghosts crucially
depends on the possibility of parameterizing the $N^i$ in terms of the
$n^i$ through (\ref{redefn}). Only then $N$, $L$ and $L_i$ appear
linearly in the action (\ref{lagbim}) and enforce the required
constraints. Introducing standard matter couplings for $\gmn$ and
$\fmn$ individually, does not change this story.\footnote{Coupling
  $\gmn$ and $\fmn$ individually to matter in the standard way, as in
  (\ref{action-gf-m_rev}), results in terms of the form (\ref{hamg}),
  now written for the metrics $g$ and $f$. Adding these to the
  bimetric action (\ref{lagbim}) simply modifies $\mathcal{C}_g$,
  $\mathcal{C}_f$ and $R^i_{gf}$, but keeps the Lagrange multipliers.
  Hence the no-ghost argument goes through unmodified.} If, instead,
one couples a combination of $g$ and $f$ to matter, one has to insure
that it does not reintroduce ghosts by destroying the constraints.

Now, we consider the matter coupling of the nonlinear massless
mode $G=g+f$ by adding (\ref{hamg}) to (\ref{lagbim}). The relevant
terms in the action are,   
\be
N\mathcal{C}_g+L\mathcal{C}_f+L_iR^i_{gf}++K\tilde{\mathcal{C}}
+K_i\tilde{\mathcal{C}}^i\,.
\label{hamg-2}
\ee
From the $3+1$ decompositions of $\Gmn$, $\gmn$ and $\fmn$ it is easy to
see that, 
\begin{align}
K_i&=N_i+L_i\,,\qquad \quad
^3\!G_{ij}=\,^3\!g_{ij}+\,^3\!f_{ij}\,,\label{shift3m}\\
K^2&=N^2+L^2+\,^3\!G^{ij}K_iK_j+-\,^3\!g_{ij}N^iN^j-\,^3\!f_{ij}L^iL^j
\,. \label{msq}
\end{align}
Already at first glance $K$ is highly nonlinear in $N$ and $L$ which
are no longer Lagrange multipliers. But this may not yet imply a
ghost. Note that after the $N^i$ have been expressed in terms of the
$n^i$ (\ref{redefn}), we may still carry out a similar
reparameterization of the $L^i$ in terms of some $l^i$. If this could
somehow render (\ref{hamg-2}) linear in $N$ and $L$, then the theory
may still have the constraints to avoid ghosts
(although it may propagate more than seven modes if the $L^i$
constraints are lost). But it turns out that $K^2$ given above is
independent of $L_i$, so reparameterizing it does not help.  

To see this, simplify the expression for $K^2$ using
(\ref{shift3m}) and writing $N^i=(N^i-L^i)+L^i$,
\be
K^2= N^2+L^2+(N^i-L^i)(N^j-L^j)
\left(^3\!g_{ij}-\,^3\!g_{ik}\,^3\!G^{kl}\,^3\!g_{lj}\right)\,.
\ee
Since $N^i-L^i=Ln^i+N{D^i}_kn^k$, in terms of $n^i$, this is
independent of $L^i$ and has the form,  
\be
K^2=c_1 N^2+c_2 L^2 + 2c_3 LN\,.
\ee
For the given $c_1$, $c_2$ and $c_3$, this expression is not a
prefect square implying that there is no way to render $K$ linear in
$N$ and $L$. Hence the associated constraints are lost. Therefore,
coupling $\Gmn$ to matter will reintroduce ghosts.
Of course, to linear order, $\delta K$ is linear in $\delta N$ and
$\delta L$ and, to this order, ghost-free matter couplings exist, as
in (\ref{Glineq}).

\subsection{Spin-2 mixing and oscillations}

In the absence of ghost-free coupling of the massless mode $\Gmn$ to
matter, one is led to consider the standard individual couplings of
$\gmn$ and $\fmn$ to matter, which are known to be ghost-free. In the
weak gravity limit, $m_g >> m_f$, we regard $\gmn$ as the
gravitational metric. The fluctuations $\dg$ and $\df$, sourced
respectively by $\delta T^g$ and $\delta T^f$, are linear combinations
of the mass eigenstates $\delta G$ and $\delta M$ given in (\ref{dGc}) and
(\ref{dMc}). So the spin-2 states are produced in the interaction
basis $(\delta g, \delta f)$ while they propagate as mass eigenstates
$(\delta M, \delta G)$. As is well known, this will lead to
oscillations between $(\delta g, \delta f)$ and a graviton $\dg$ may
oscillate to the other spin-2 field $\df$. This is very similar to
neutrino oscillations or the $K_0-\bar K_0$ oscillations. The
detectability of this effect reduces for higher FP mass of the 
massive mode. So in cases where the massive spin-2 state can be
interpreted as a meson or a heavy elementary particle, the effect is
negligible. But it will have consequences for very light spin-2
states. Nevertheless it remains an interesting consequence of the
inconsistency of coupling the massless field to matter that in the
presence of a neutral massive spin-2 field, the gravitational force is 
mediated by a particle that is a superposition of mass eigenstates.

\section{Action for the nonlinear massive spin-2 field}\label{sec_nonlin_M}

In this section we consider the bimetric action (\ref{action}) as a
theory of a massive spin-2 field $\Mmn$ in the presence of a
gravitational metric $\gmn$. We also obtain the ghost-free couplings
of $\Mmn$ to fermionic matter fields.

\subsection{The action in terms of \texorpdfstring{$g$}{g} and 
\texorpdfstring{$M$}{M}}

To regard the bimetric action (\ref{action}) as a theory of a massive
spin-2 field in the presence of gravity, we express it in terms of the
nonlinear massive field $\Mmn$ and the metric $\gmn$.\footnote{Other
  possibilities would be to write the bimetric action in terms of the
  nonlinear massless field $\Gmn$ (\ref{Gnl}) and the massive field
  $M$ or $M^G$. The $G-M^G$ action is given in the appendix. In the
  $G-M$ case the expressions for $g$ and $f$ become too involved.
  Since $\Gmn$ cannot be coupled to matter in a ghost-free way, here
  we concentrate on the $g-M$ case.} $\gmn$ couples to matter in the
standard way and is the gravitational metric. $\Mmn$ is a massive
spin-2 field and couples non-minimally to gravity. The two spin-2
fields mix and their mass and interaction eigenstates do not coincide,
just as for spin-$\half$ fields in the standard model. These mixings
also result in deviations from GR. In section 3, in the linearized
theory, the mixings became small in the weak gravity limit
$m_g>>cm_f$. The hope is that in this limit the mixings remain small
even nonlinearly and the predictions of this theory do not greatly
differ from GR.

To express the bimetric action (\ref{action}) in terms of the fields
$g$ and the massive spin-2 field $M$, let us start with the potential
$V(S\,,\beta_n)$, where $S=\sqrt{g^{-1}f}$ (\ref{int_pot}), and use,
\be
S^\rho_{\,\,\nu}= g^{\rho\sigma}M_{\sigma\nu}+c\delta^\rho_{\,\,\nu}.
\label{SM}
\ee
Then, the potential written in terms of elementary symmetric
polynomials, becomes \cite{Hassan:2011vm},  
\be
V(S,\beta_n)=\sum_{n=0}^4\beta_n e_n(S)=\sum_{n=0}^4\beta_n\,c^n 
e_n({\mathbb 1}+g^{-1}M/c)=\sum_{n=0}^4\alpha_n^c e_n(g^{-1}M)=
V(g^{-1}M, \alpha_n^c)
\label{VM}
\ee
The last step follows from the linear relations between the
$e_n(\mathbb X)$ and $e_n({\mathbb 1}+{\mathbb X})$.\footnote
{Explicitly,  
$e_0(\mathbb{1+X})=e_0(\mathbb{X})$, 
$e_1(\mathbb{1+X})=4e_0(\mathbb{X})+e_1(\mathbb{X})$, 
$e_2(\mathbb{1+X})=6e_0(\mathbb{X})+3e_1(\mathbb{X})+e_2(\mathbb{X})$, 
$e_3(\mathbb{1+X})=4e_0(\mathbb{X})+3e_1(\mathbb{X})+2 e_2(\mathbb{X})
+ e_3(\mathbb{X})$, and 
$e_4(\mathbb{1+X})=e_0(\mathbb{X})+e_1(\mathbb{X})+e_2(\mathbb{X})
+e_3(\mathbb{X}) + e_4(\mathbb{X}).$ }
The parameters $\alpha_n^c$ are given in terms of $\beta_n$ as,
\begin{align}
\alpha_4^c&=\beta_4\,,\qquad\quad
\alpha_3^c=\beta_3+c\beta_4\,,\qquad
\alpha_2^c=\beta_2+2c\beta_3+c^2\beta_4\,,\nn\\
\alpha_1^c&=\beta_1+3c\beta_2+3c^2\beta_3+c^3\beta_4\,,\qquad
\alpha_0^c=\beta_0+4c\beta_1+6c^2\beta_2+4c^3\beta_3+c^4\beta_4\,.\nn
\end{align}
An advantage of writing the theory in terms of $M$, of course, is
that the potential no longer involves a square-root matrix. 
Also, this form is analogous to the familiar form of mass terms in
field theory, for example, the mass term for massive vector fields,
$\sqrt{-g}\, g^{\mu\nu} A_\mu A_\nu$. 

Now let us turn to the kinetic term for the massive field $M$ which is
obtained from $\sqrt{-f}R(f)$ on expressing $f$ in terms of $M$ and $g$. 
The criterion is that in the final action for $\Mmn$, all covariant
derivatives must be with respect to the metric $\gmn$. To achieve this
in a systematic way, it is convenient to use, 
\be
\fmn=g_{\mu\rho} (S^2)^\rho_{\,\,\nu} \,,
\label{fS}
\ee
where $S$ is related to $M$ in a simple way (\ref{SM}). Now, the
curvatures of $f$ can be expressed in terms of curvatures of $g$ using
the results in appendix \ref{app_Rrels}. In particular, (\ref{RNabla})
gives, 
\be
R_{\mu\nu}(f)=R_{\mu\nu}(g)+
2\nabla_{[\mu}C_{\alpha]\nu}^{\ph\alpha\ph]\ph\nu\alpha}
-2C_{\nu[\mu}^{\ph\nu\ph[\ph\mu\beta}C_{\alpha]\beta}^{\ph\alpha\ph]\ph\beta\alpha}\,,
\label{Rgf}
\ee
where $\nabla$ is the covariant derivative compatible with $\gmn$ and,
\be
C_{\mu\nu}^{\ph\mu\ph\nu\alpha}=\tfrac1{2}\phi^{\alpha\beta}\left(
\nabla_\mu f_{\beta\nu}+\nabla_\nu f_{\beta\mu}
-\nabla_\beta f_{\mu\nu}\right)\,.
\ee
Here, $f$ is given by (\ref{fS}) and, for ease of notation we have
introduced,
\be
\phi^{\mu\nu}\equiv (f^{-1})^{\mu\nu}=(S^{-2})^\mu_{~\rho}\,g^{\rho\nu}\,.
\label{phi_def}
\ee
Using the curvature relation above, along with $\sqrt{-f}=\sqrt{-g}
\det(S)$ and $R(f)=\phi^{\mu\nu}R_{\mu\nu}(f)$, it is a
straightforward though tedious exercise to show that (modulo total
derivatives), 
\be
\sqrt{-f}R(f)=\sqrt{-g}\det(S)\,\left[\phi^{\mu\nu}R_{\mu\nu}(g)+
\Pi^{\sigma\pi\alpha\beta}_{\rho\omega}
\nabla_\alpha S^\rho_{~\sigma}\nabla_\beta S^\omega_{~\pi}\right]\,.
\label{RfRg}
\ee
The ``polarization'' tensor $\Pi$ is a function of $g$ and $M$
(through $S$) given by, 
\be
\Pi_{\rho\omega}^{\sigma\pi\alpha\beta}
=\biggl(2\phi^{\alpha\mu}\phi^{\kappa\delta}\phi^{\beta\nu}
-2\phi^{\alpha\mu}\phi^{\beta\kappa}\phi^{\delta\nu}
-\phi^{\alpha\beta}\phi^{\kappa\delta}\phi^{\mu\nu}
+\phi^{\alpha\beta}\phi^{\kappa\mu}\phi^{\delta\nu}\biggr)
\delta^\sigma_{(\kappa}\delta^\gamma_{\delta)}\,  
\delta^\pi_{(\mu} \delta^\lambda_{\nu)}\, S_{\gamma\rho}\,
S_{\lambda\omega}. 
\label{poltensor_g}
\ee
The discarded total derivative terms in (\ref{RfRg}) arise since the
left-hand side has $f\p^2 f$ terms, leading to $M\p^2 M$ terms
on the right-hand side. These have been converted to $\p M\p M$
terms which amounts to adding the Gibbons-Hawking-York boundary term to
the action for $f$. 

Finally, putting all this together, the action for a massive spin-2
field $M$ interacting with a gravitational metric $g$ is given by,
\begin{align}
\label{action_gM}
S_{gM} =\int\td^4x\sqrt{-g}~\Big[&
m_g^2R(g) +m_f^2\det(S)\,\phi^{\mu\nu}R_{\mu\nu}(g)  \nonumber \\
& + m_f^2\det(S)\, \Pi_{\rho\omega}^{\sigma\pi\alpha\beta}\nabla_\alpha 
M^\rho_{\ph\rho\sigma}\nabla_\beta M^\omega_{\ph\omega\pi}
-2m^4\,V(g^{-1}M, \alpha_n^c) \Big] \,.
\end{align}
Here $S^\mu_{\,\,\nu}$ is a function of $\Mmn$. The coupling of $M$ to
matter fields will be discussed below. 

The perturbative content of the $(g,M)$ action (\ref{action}) can be 
discerned easily. The second order action for the fluctuations, 
$g=\bar g+\delta g$ and $M=0+\delta M$, will contain $(\delta g)^2$,
$\delta g\delta M$, and $(\delta M)^2$ terms. It can be diagonalized
in terms of the massless mode $\delta G$ and massive mode
$\delta M$, leading to the linearized equations (\ref{Glineq}) and
(\ref{Mlineq}).  

\subsection{Some features of the \texorpdfstring{$g-M$}{g-M} action}

{\it Equivalence of the two formulations}: It is straightforward that
the $g$ and $M$ equations of motion obtained from (\ref{action_gM})
imply the bimetric $g$ and $f$ equations of motion and vice versa,
\be
\frac{\delta S_{gf}}{\delta f}\big|_{g}=
\frac{\delta S_{gM}}{\delta M}\big|_{g}
\frac{\delta M}{\delta f}  =0\,, 
\qquad
\frac{\delta S_{gf}}{\delta g}\big|_{f}=
\frac{\delta S_{gM}}{\delta g}\big|_{M}
+\frac{\delta S_{gM}}{\delta M}\big|_{g}\frac{\delta M}{\delta g}=0\,.
\ee
Hence the $g-M$ formulation is classically equivalent to the $g-f$
formulation and, in particular, is also ghost-free. The $g-M$ form has
the advantage that it does not involve square-root matrices. The price
one pays on the other hand, is the tedious kinetic structure for $M$
and its kinetic mixing with gravity. Although deriving the equations
of motion from (\ref{action_gM}) is not convenient, it is much easier
to obtain these equations by starting with the $g-f$ equations
(\ref{g_eom}) and (\ref{f_eom}), eliminating $R(f)$, and then converting
$R_{\mu\nu}(f)$ to $R_{\mu\nu}(g)$ using (\ref{Rgf}).

In the $g-M$ action, one may perform perturbative calculations even
around non-vanishing $M$ backgrounds. In the $g-f$ formulation,
performing higher order perturbative calculations around
non-proportional backgrounds is not straightforward as in that case
expanding $\sqrt{g^{-1}f}$ is not simple. Of course, the two
formulations are not expected to be equivalent in quantum theory
unless one takes into account the Jacobian factor that arises from the
change of variables.

{\it Parameterizing deviations from general relativity}: Most of the
classical solutions of the bimetric action $S_{gf}$ do not coincide
with classical solutions in general relativity
\cite{Volkov:2011an,Volkov:2012wp,vonStrauss:2011mq,Comelli:2011wq,
  Comelli:2011zm,Baccetti:2012ge,Deffayet:2011rh}, except for the
class of proportional backgrounds $\bar f=c^2\bar g$ considered here,
with $c$ determined by the parameters of the theory. Generic matter
couplings of the $g$ and $f$ metrics will drive the solutions away
from proportional backgrounds. We are interested in parameter regions
where these deviations are small.

In the $g-M$ formulation, proportional backgrounds correspond to $\bar
M_{\mu\nu}=0$. Hence, in the field theory language, a vanishing vacuum
expectation value for the massive spin-2 field implies that the
classical solutions for the metric coincides with general
relativity. Hence, deviations of the bimetric theory from general
relativity are parameterized by the deviations of $\Mmn$ from
zero. These are driven by general couplings of $\gmn$ and $\Mmn$ to
matter fields that violate the condition $\bar T^f=\alpha^2\bar T^g$
(\ref{constcosconst}). 

{\it Energy-momentum tensor of spin-2 fields} : Consider the
gravitational energy momentum tensor $\delta S_{gf}/\delta g_{\mu\nu}$. 
In the $g-f$ formulation, $\fmn$ contributes to this only through the
potential $V$, but its kinetic term $\sqrt{-f}R(f)$ does not
gravitate (of course, it still affects the dynamics of $\gmn$ since
their equations of motion are coupled). This is while the Noether
energy momentum tensor computed around flat $\fmn$, will receive
contributions from $\sqrt{-f}R(f)$. 

In the $g-M$ formulation, the gravitational energy momentum tensor
$\delta S_{gM}/\delta g_{\mu\nu}|_M$ contains contributions from the
kinetic term of $M$, as well from $V$ and these appear consistent with
the contribution to the Noether energy momentum tensor around flat
$g$. Of course the complete set of equations is the same in both
formalisms as these contributions drop out on imposing the $M$
equation of motion. The same statements apply to the matter couplings
of $f$, that is $S(f,\psi)$. In the $g-M$ formulation, $M$ is mostly
minimally coupled to $g$ (in the sense that the curved space form can
be constructed from the flat space expression) except for the
non-minimal $\phi^{\mu\nu}R_{\mu\nu}$ term.

{\it Comparison to earlier work} : The $g-M$ action (\ref{action_gM})
is useful in comparing to earlier attempts of writing a theory of
massive spin-2 on a gravitational background. For example, the
approach in \cite{hep-th/9908028, hep-th/9910188} was to start with
the quadratic FP theory in flat spacetime and covariantize it with a
metric $\gmn$, also adding non-minimal curvature couplings. 
This procedure will in general not reproduce the action
(\ref{action_gM}) as it will miss the factor $\det(S)$ as well as the
complicated polarization structure (\ref{poltensor_g}), since it only
considers terms quadratic in the massive field. 

\subsection{Coupling massive spin-2 fields to matter}

At present, the only known ghost-free matter couplings in bimetric
theory are the standard couplings of $\gmn$ and $\fmn$ to matter
sources, as in (\ref{action-gf-m_rev}). In the weak gravity limit, we
interpreted the $\gmn$ couplings as the gravitational interactions of
matter fields just as in GR. Then the $\fmn$ couplings give rise to
very specific interactions of the massive spin-2 field with matter,
dictated by the absence of ghost. To write $f$ in terms of $M$, again
it is convenient to proceed through the related matrix
$S=g^{-1}M-c{\mathbb 1}$ (\ref{SM}) and use,
\be
\fmn = S_\mu^{~\alpha}g_{\alpha\beta}S^\beta_{~\nu} = 
g_{\mu\alpha}S^\alpha_{~\beta}S^\beta_{~\nu}\,.\qquad
\label{fSgS2}
\ee
For $\fmn$ couplings to bosonic matter, the manipulations are
straightforward. For example, for a Proca field, 
$\Lag(f,A)=-\tfrac1{4}\sqrt{f}\left[f^{\mu\nu}f^{\kappa\lambda}
F_{\mu\kappa}F_{\nu\lambda} + 2m_A^2f^{\mu\nu}A_\mu A_{\nu}\right]\,,$
where $F_{\mu\nu}=\p_\mu A_\nu-\p_\nu A_\mu$, One obtains
the coupling of $A_\mu$ to the massive spin-2 field by expressing $f$ 
in terms of $M$ and $g$ through $S$. Then, on raising some indices
using $g^{\mu\nu}$ one gets,
\be
\Lag(g, M, A)=-\tfrac1{4}\sqrt{g}\det S\,\bigl[(S^{-2})^\mu_{~\rho}
(S^{-2})^\kappa_{~\sigma}F_{\mu\kappa}F^{\rho\sigma}
+2m_A^2(S^{-2})^\mu_{~\rho}A_\mu A^{\rho}\bigr]\,.
\ee
The details of this Lagrangian can be investigated further by, 
e.g., considering flat space $g=\eta$, and/or expanding 
$S^\mu_{~\nu}=c\delta^\mu_{\nu}+a\bar g^{\mu\lambda}\delta
M_{\lambda\nu}$.

For fermionic matter a little more work is needed since fermions
couple to spin-2 fields through vielbeins. For the $\fmn$ and $\gmn$
vielbeins (below we use the conventions of 
\cite{Ortin:2004ms}), 
\be
\fmn =\tilde e_\mu^{~a}\eta_{ab}\tilde e^b_{~\nu}\,,\quad \gmn =
e_\mu^{~a}\eta_{ab} e^b_{~\nu}\,, 
\label{vierbein}
\ee
equation (\ref{fSgS2}) implies the relation,
\be
\tilde e^a_{~\mu} = \Lambda^a_{~b} e^b_{~\nu} S^\nu_{~\mu}\,.
\label{efegM}
\ee
Here $\Lambda$ is an arbitrary Lorentz transformation,
$\Lambda^{\mathrm{T}}\eta\Lambda=\eta$. In a Lorentz invariant 
theory $\Lambda$ drops out of all expressions, so we set
$\Lambda=\mathbb{1}$  without loss of generality. 
The vielbeins enter the couplings through the curved space
$\gamma$-matrices and through the spin-connections. In terms of the
Lorentz frame $\gamma$-matrices $\bar\gamma^a$, one constructs a pair
of curved space $\gamma$-matrices $\tilde\gamma^\mu=\tilde e^\mu_{~a}
\bar\gamma^a$ and $\gamma^\mu=e^\mu_{~a}\bar\gamma^a$. They satisfy,
\be
\{\bar\gamma^a,\bar\gamma^b\} = 2\eta^{ab}\,,\qquad
\{\tilde\gamma^\mu,\tilde\gamma^\nu\} = 2f^{\mu\nu}\,,\quad
\{\gamma^\mu,\gamma^\nu\} = 2g^{\mu\nu}\,.
\label{gammadef}
\ee
The inverse of the relation (\ref{fSgS2}), with $\Lambda=1$, then 
implies,\footnote{A general $\Lambda$ is absorbed by a Lorentz
  transformation of the spinors, $\psi'=A\psi$ where, 
  $\Lambda^a_{~b}\bar\gamma^b=A^\dagger\bar\gamma^aA$.} 
\be
\tilde\gamma^\mu = (S^{-1})^\mu_{~\nu}\gamma^\nu\,.
\ee
Fermions also couple to vielbeins through spin connections in Lorentz
covariant derivatives acting on them, 
\be
\tilde D_\mu = \p_\mu -\tfrac1{8}\tilde w_\mu^{~ab}[\bar\gamma_a,
\bar\gamma_b]\,. 
\ee
The spin-connection is given in terms of vielbeins and the Christoffel
connection through,
\be
\tilde w_\mu^{~ab} = \tilde e^b_{~\nu}\p_\mu[\eta^{ac}\tilde e_c^{~\nu}] 
+ \tilde e^b_{~\sigma}\eta^{ac}\tilde e_c^{~\nu}\tilde
\Gamma_{\mu\nu}^{\ph\mu\ph\nu\sigma}\,. 
\label{wdef}
\ee
Using (\ref{efegM}), with $\Lambda=1$, we can rewrite this as,
\be
\tilde w_\mu^{~ab} = \left(e^b_{~\rho}S^\rho_{~\nu}
\p_\mu[(e^{a\alpha}(S^{-1})_\alpha^{~\nu}] 
+ e^b_{~\lambda}S^\lambda_{~\sigma}
e^{a\alpha}(S^{-1})_\alpha^{~\nu}
\tilde\Gamma_{\mu\nu}^{\ph\mu\ph\nu\sigma}\right)\,.
\label{wlambda}
\ee
From appendix \ref{app_Rrels}, $\tilde\Gamma$ is related to the
Christoffel connection $\Gamma$ of $\gmn$ by,
\be
\tilde\Gamma_{\mu\nu}^{\ph\mu\ph\nu\sigma}=\Gamma_{\mu\nu}^{\ph\mu\ph\nu\sigma}
+C_{\mu\nu}^{\ph\mu\ph\nu\sigma}\,,
\qquad
C_{\mu\nu}^{\ph\mu\ph\nu\sigma} = \tfrac1{2}f^{\sigma\rho}\left(
\nabla_\mu f_{\rho\nu}+\nabla_\nu f_{\mu\rho}-\nabla_\rho
f_{\mu\nu}\right)\,, 
\ee
where the covariant derivatives are with respect to $\gmn$, and where
$\fmn$ is regarded as a function of $g$ and $M$ through (\ref{fSgS2}).
Using these relations, it is straightforward to re-express any
coupling of $\fmn$ to fermions in terms the massive spin-2 field  
$\Mmn$ and the gravitational metric $\gmn$. The resulting expressions
are highly nonlinear in the fields.

As an example, consider the couplings to lowest order in the
fluctuation $\delta M$ of the massive spin-2 field around the $\bar
M=0$ background. Then, $S^\mu_{~\nu}=c\delta^\mu_\nu+a\bar
g^{\mu\lambda}\delta M_{\lambda\nu}$, and to first order,
\be
C_{\mu\nu}^{\ph\mu\ph\nu\sigma}=\tfrac{a}{c}g^{\sigma\rho}
\left[2\nabla_{(\mu}\delta M_{\nu)\rho}
-\nabla_\rho\delta M_{\mu\nu}\right]\,.
\ee
Similarly, to this order,
\be
\tilde\gamma^\mu = \tfrac1{c}\gamma^\mu -\tfrac{a}{c^2}
\bar g^{\mu\lambda}\delta M_{\lambda\nu} \gamma^\nu\,, \qquad
\tilde D_\mu = D_\mu -\tfrac{a}{4c}[\gamma^\rho,\gamma^\sigma]
\nabla_{[\rho}\delta M_{\sigma]\mu}\,.
\label{gamma_dM}
\ee
Let us apply these to the coupling of $\fmn$ to a spin-$\tfrac{1}{2}$
field $\psi$,
\be
\mathcal{L}_{1/2}=i\sqrt{-f}~\bar{\psi}(\tilde\gamma^\mu\tilde{\mathcal
  D}_\mu+im_\psi)\psi + \mathrm{h.c.}\,,
\label{lag_ferm}
\ee
in which $\bar\psi=\psi^\dagger\bar\gamma^0$. $\tilde{\mathcal{D}}_\mu
=\tilde D_\mu+iqA_\mu$ is the Lorentz and gauge covariant derivative
with Abelian gauge field $A_\mu$. We write this to linear order in
$\delta M$ and in the flat space limit $\gmn=\eta_{\mu\nu}$. Using, 
$\sqrt{-f}=c^4+c^3a\delta M^\rho_{~\rho}+\cdots$, one has,
\begin{align}
\mathcal{L}_{1/2}=&c^3\left(1+\tfrac1{c\meff}\delta M^\rho_{~\rho}
\right)\Lag_{\mathrm{free}} \nn\\
&\qquad
-i\tfrac{c^3}{c\meff}~\bar\psi\left(\delta M^\mu_{~\nu}\bar\gamma^\nu\p_\mu
+iq\delta M^\mu_{\,\,\nu}\bar\gamma^\nu A_\mu+\tfrac1{4}\bar\gamma^\mu
[\bar\gamma^\rho,\bar\gamma^\nu]\p_{[\rho}\delta M_{\nu]\mu}\right)\psi
	+\mathrm{h.c.}\,,
\label{lag_fermflat}
\end{align}
where $\Lag_{\mathrm{free}}=i~\bar{\psi}\left(\gamma^\mu \p_\mu+iq
\bar\gamma^\mu A_\mu+icm_\psi\right)\psi$. After a partial integration
and using (\ref{gammadef}), the derivative couplings become,
\begin{align}
-\frac{ic^3}{c\meff}\delta M_{\mu\nu}\Big[\tfrac{3}{8}\bar\psi
(\gamma^\mu\p^\nu+\gamma^\nu\p^\mu)\psi - & \tfrac1{8}
(\p^\mu\bar\psi\gamma^\nu+\p^\nu\bar\psi\gamma^\mu)\psi
\nn\\
+ &\frac1{4}\eta^{\mu\nu}\p_\rho\bar\psi\gamma^\rho\psi-\frac{3}{4}
\eta^{\mu\nu}\bar\psi\gamma^\rho\p_\rho\psi\Big]+\mathrm{h.c.}\,.
\end{align}
Finally, considering the hermitian conjugate (with the usual
hermiticity condition $(\gamma^\mu)^\dagger=\gamma^0\gamma^\mu
\gamma^0$), this can  be written,  
\be
-\frac{ic^3}{c\meff}\delta M_{\mu\nu}\left[\frac{1}{2}
\bar\psi(\gamma^\mu\p^\nu+\gamma^\nu\p^\mu)\psi
-\eta^{\mu\nu}\bar\psi\gamma^\rho\p_\rho\psi\right]+\mathrm{h.c.}\,.
\label{dMderpsi}
\ee
Couplings of this form were recently considered in a phenomenological
context in \cite{Grinstein:2012pn}, to address the top-quark
forward-backward asymmetry. Here, in contrast to
\cite{Grinstein:2012pn}, the couplings are not {\it a priori} expected  
to be flavor violating since they come only from the Lorentz covariant
derivative and are essentially of a purely gravitational nature. In
particular, the first term of (\ref{dMderpsi}) is simply the 
stress-energy tensor while the second corresponds to a non-derivative 
trace coupling on-shell. 

\section{Generalization to more than one massive field} 

Recently, in \cite{Hinterbichler:2012cn} the bimetric action was
generalized to a ghost-free theory of $\mathcal{N}$ interacting spin-2
fields. In this theory, the kinetic term is given in terms of
$\mathcal{N}$ metrics $\gmn(I)$ and the interactions between these
are constructed in terms of the corresponding vielbeins
$e^a_{\,\,\mu}(I)$,   
\be\label{multi-e}
\sum_{I=1}^{\mathcal{N}}\int\mathrm{d}^4x\sqrt{-g(I)}~R(I)+
\frac{m^2}{4}\int \md^4x\, U[e(1),\cdots, e(\mathcal{N})]\,.
\ee
The potential $U$, constructed in \cite{Hinterbichler:2012cn}, will be
presented below in a reformulation. Around flat backgrounds, 
the spectrum consists of one massless and $\mathcal{N}-1$ massive
states. The vielbein description is elegant and was very convenient
for showing the absence of the Boulware-Deser ghosts. For further
work, see \cite{Hassan:2012wt}.

Here we are interested in interpreting (\ref{multi-e}) as a theory of
$\mathcal{N}-1$ spin-2 fields in the presence of gravity. First one
has to identify one of the vielbeins, say, $e^b_{\,\,\nu}(1)$, with
the gravitational metric, $\gmn=e_\mu^{\,\,a}(1)\eta_{ab}
e^b_{\,\,\nu}(1)$. Then, off shell, the $16(\mathcal{N}-1)$ components
of the remaining vielbeins contain the $10(\mathcal{N}-1)$ degrees of
freedom for describing $\mathcal{N}-1$ spin-2 fields as symmetric
rank-2 tensors with kinetic terms consistent with general covariance.
In addition, there are $6({\mathcal N}-1)$ extra non-dynamical fields,
as there are no leftover local Lorentz transformations to remove
them. The latter, have to be eliminated through their equations of
motion to isolate the spin-2 content of the theory. A difficulty that
arises for $\mathcal{N}>2$ is in disentangling these non-dynamical
components from the ones belonging to the spin-2 fields in kinetic
terms \cite{Hinterbichler:2012cn}. 

In other words, from the remaining vielbeins, one can construct
$\mathcal{N}-1$ rank-2 tensors $\theta^\mu_{\,\,\nu}(I)=
e^\mu_{\,\,a}(1) \,e^a_{\,\,\nu}(I)$ of mixed symmetry. The potential
is a function of the $\theta^\mu_{\,\,\nu}(I)$. It is difficult to
extract from these the spin-2 fields that have kinetic terms, by
solving the non-dynamical equations. Even more difficult is doing so
in a general covariant way. 

This issue is addressed in the metric formulation of the multivielbein
action (\ref{multi-e}) that was obtained, and argued to remain
ghost-free, in \cite{Hassan:2012wt}. In this setup, the non-dynamical
fields are isolated from the spin-2 content in a generally covariant
way without solving any equations of motion, making it appropriate for
the considerations here. Then we work with the multi spin-2 action,
\be
\sum_{I=1}^{\mathcal{N}}\int\mathrm{d}^4x\sqrt{-g(I)}~R(I)+
\frac{m^2}{4}\int \md^4x\,T^{I_1\hdots I_4}\,U_{I_1\hdots I_4 } \,,
\label{multi-g}
\ee
where $T_{I_1I_2I_3I_4}$, totally symmetric in its indices, contains
the free parameters of the theory and the multivielbein potential of 
\cite{Hinterbichler:2012cn} is reformulated to \cite{Hassan:2012wt}, 
\begin{align}
U_{I_1\cdots I_4} = & \sqrt{-\det{g(1)}}\,\,
\te^{\mu_1\cdots \mu_4}\,\te_{\nu_1\cdots \nu_4}\,  \nn\\  
& \qquad\qquad\times\, 
L^{\nu_1}_{~\lambda_1}(I_1)\Big[\sqrt{\ds g^{-1}(1)\,g(I_1)}\,
\Big]^{\lambda_1}_{~\mu_1}\,\cdots 
L^{\nu_4}_{~\lambda_4}(I_4) 
\Big[\sqrt{\ds g^{-1}(1)\,g(I_4)}\,\Big]^{\lambda_4}_{~\mu_4}\,.  
\label{massMetric}
\end{align}
In this expression the $L^{\nu}_{~\lambda}(I)$ satisfy $\gmn
L^{\mu}_{~\rho}(I)L^{\nu}_{~\sigma}(I)= g_{\rho\sigma}$ and
$L^{\nu}_{~\lambda}(1)=\delta^{\nu}_{~\lambda}$. They carry the
$6({\mathcal N}-1)$ non-dynamical parameters as they do not enter the
kinetic terms. They can be eliminated through their own equations of
motion, but solving these equations is not necessary to identify the
spin-2 content of the theory. In this form, the similarity to bimetric
form as given in \cite{Hassan:2011vm} is apparent. 

The equations of motion obtained from (\ref{multi-e}) or
(\ref{multi-g}) admit proportional background solutions,  
\be
\bgmn(I)=c^2_I\,\bgmn \,,\qquad {\rm or}\qquad  
\bar e^a_{\,\,\mu}(I)=c_I\,\bar e^a_{\,\,\mu}\,,
\ee
for $I=2,\cdots,\mathcal{N}$, and were we have denoted $\gmn(1)\equiv 
\gmn$ and $e^a_{\,\,\mu}(1)\equiv\bar e^a_{\,\,\mu}$. The $c_I$ will
be determined by the parameters of theory. In analogy with
the bimetric case, one can introduce, for $I=2,\cdots,\mathcal{N}$,  
\be\label{defmi}
S(I)\equiv\sqrt{\ds g^{-1}\,g(I)}\,,\qquad 
M(I)=g\,S(I)-c_I g\,.
\ee
Since, 
\be
g\,S(I)=[\,g\,S(I)\,]^{\mathrm{T}}\,,
\ee
the $\Mmn(I)$ will be symmetric and represent the $\mathcal{N}-1$
massive spin-2 fields with vanishing expectation values in proportional
backgrounds. These generalize the massive mode (\ref{Mnl}) of the
bimetric theory in the picture that $\gmn$ is the gravitational
metric. Note however that for generic coefficients $T_{I_1I_2I_3I_4}$
in (\ref{multi-g}), the actual mass eigenstates will be given by
linear combinations of the $\Mmn(I)$.

In terms of $g$ and $M(I)=gS(I)-c_Ig$, the potential
$\sqrt{-g}~V\big(L(I)S{(I)}\big)$ is a finite polynomial of its
argument. The kinetic terms for the $M(I)$ will simply involve
$\mathcal{N}-1$ copies of the corresponding terms in the bimetric
case. The $L(I)$ are determined in terms of $g$ and the $M(I)$.

\section{Discussion}\label{sec_conclusion}

The results have already been summarized in section 1 so here we only
make some additional comments. The nonlinear massless and massive
modes were introduced as an extension of the corresponding linear
modes. It remains to be seen if they have a relevance directly at the
nonlinear level. Although it is stated that the weak gravity limit is
needed to approach GR solution in a generic sense, and that the
non-vanishing VEV of the massive mode $M$ parameterizes deviations
from GR, these effects have not yet been quantified. For example, note
that the Bianchi constraints (\ref{Bianchi}) are independent of $m_g$
and $m_f$ hence their nontrivial consequences will not be affected by
the weak gravity limit.

Another feature of the bimetric theory is that in the $g-f$
formulation with $g$ as the gravitational metric, the kinetic energy
of $f$ as well as its matter couplings affect gravity only through the
potential $V(g^{-1}f)$. In the $g-M$ formulation there are direct
couplings between $g$ and the kinetic term as well as matter
interactions of $M$. However, after the $M$ equation of motion is
imposed, the two sectors interact only through $V$ again. In this
sense the couplings in spin-2 theories are maximally non-minimal.  

\vspace{20pt}

\acknowledgments

We would like to thank Jonas Enander, Paolo Gondolo, Alexander Merle,  
Stefan Sj\"{o}rs, Rachel Rosen and Bo Sundborg for useful discussions
and comments.

\vspace{.5cm}

\appendix


\section{Curvature relations}
\label{app_Rrels}
Here we provide a relation between Ricci tensors on a manifold 
endowed with two covariant derivatives (see, for example,  
\cite{Wald:1984rg}). This simplifies manipulation in bimetric 
theory.  

{\bf General relations:}
Given any two derivative operators $\nabla$ and $\bar\nabla$, there
exist a ($1,2$) tensor field $C$ such that the actions on
vectors $\omega_\mu$ are related by,
\be
\nabla_\mu\omega_\nu = \bar\nabla_\mu\omega_\nu
-C_{\mu\nu}^{\ph\mu\ph\nu\alpha}\omega_\alpha\,.
\ee
If $\nabla$ and $\bar\nabla$ are torsion free and compatible with
metrics $g$ and $\bar g$, the tensor $C$ is given by, 
\be\label{Cg}
C_{\mu\nu}^{\ph\mu\ph\nu\alpha}=\tfrac1{2}g^{\alpha\beta}\left(
\bar\nabla_\mu g_{\beta\nu}+\bar\nabla_\nu g_{\beta\mu}
-\bar\nabla_\beta \gmn\right)\,.
\ee
Defining the associated Riemann tensors by 
$\left[\nabla_\mu,\nabla_\alpha\right]\omega_\nu =
-R_{\mu\alpha\nu}^{\ph\mu\ph\alpha\ph\nu\beta}\omega_\beta$, it is
straightforward to derive a relation between the Ricci tensors  
$R_{\mu\nu}=R_{\mu\alpha\nu}^{\ph\mu\ph\alpha\ph\nu\alpha}$ as, 
\be\label{RNabla}
R_{\mu\nu}(g) = R_{\mu\nu}(\bar g) 
+ 2\bar \nabla_{[\mu}C_{\alpha]\nu}^{\ph\alpha\ph]\ph\nu\alpha}
-2C_{\nu[\mu}^{\ph\nu\ph[\ph\mu\beta}C_{\alpha]\beta}^{\ph\alpha\ph]\ph\beta\alpha}\,.
\ee

{\bf Example: Linearizing General Relativity:}
Consider a metric $g$ as a perturbation around a background metric
$\bar g$, $\gmn = \bgmn + \delta g_{\mu\nu}$\,.
To linear order in $\delta g$, the curvature relation (\ref{RNabla})
gives,
\be\label{RNablagr}
R_{\mu\nu}(g) = R_{\mu\nu}(\bar g) 
+ 2\bar \nabla_{[\mu}\delta\Gamma_{\alpha]\nu}^{\ph\alpha\ph]\ph\nu\alpha}\,,
\ee
where, with an obvious change of notation, $\delta\Gamma$ is given by
the linear terms in (\ref{Cg}),  
\be
\delta\Gamma_{\mu\nu}^{\ph\mu\ph\nu\alpha}=\tfrac1{2}\bar g^{\alpha\beta}\left(
\bar\nabla_\mu\delta g_{\beta\nu}+\bar\nabla_\nu\delta g_{\beta\mu}
-\bar\nabla_\beta\delta\gmn\right)\,.
\ee
This can be used to expand the Einstein equations,
$R_{\mu\nu} -\tfrac{1}{2}g_{\mu\nu} R+\Lambda g_{\mu\nu}=
\frac1{M_P^2}T_{\mu\nu}$, to linear order in $\delta g$. One gets the 
background and the fluctuation equations,
\be
\bar R_{\mu\nu}-\frac{1}{2}\bgmn\bar R +\Lambda \bgmn
= \frac{1}{M_P^2}\bar T_{\mu\nu}\,,
\hspace{1.5cm}
\bar{\mathcal{E}}_{\mu\nu}^{\rho\sigma} \,
\delta g_{\rho\sigma} +\Lambda \delta g_{\mu\nu}
= \frac{1}{M_P^2}\delta T_{\mu\nu}\,,
\ee
where we have defined, 
\begin{align}
\label{A_ERdef}
\bar{\mathcal{E}}^{\rho\sigma}_{\mu\nu}\delta g_{\rho\sigma} 
&=-\tfrac1{2}\Big[\delta^\rho_\mu\delta^\sigma_\nu\bar\nabla^2
+\bar g^{\rho\sigma}\bar\nabla_\mu\bar\nabla_\nu -\delta^\rho_\mu
\bar\nabla^\sigma\bar\nabla_\nu-\delta^\rho_\nu\bar\nabla^\sigma
\bar\nabla_\mu
\nn\\
&\hspace{3cm} 
-\bar g_{\mu\nu}\bar g^{\rho\sigma}\bar\nabla^2 
+\bar g_{\mu\nu}\bar\nabla^\rho\bar\nabla^\sigma
-\bar g_{\mu\nu}\bar R^{\rho\sigma}
+\delta^\rho_\mu \delta^\sigma_\nu\bar R\Big]\delta g_{\rho\sigma}\,.
\end{align}
Using the background equation, the curvature contributions to 
$\bar{\mathcal{E}}_{\mu\nu}^{\rho\sigma}$ can be re-expressed in terms
of $\bar T_{\mu\nu}$ and $\Lambda$.

\section{Bimetric action in scaled variables}
The action (\ref{action}) can be recast in a more symmetric form.
Consider the rescalings,
\be\label{A_variables_scaled}
g_{\mu\nu}=\frac{\meff^2}{m_g^2}\,\tilde g_{\mu\nu}\,,\qquad
f_{\mu\nu}=\frac{\meff^2}{m_f^2}\,\tilde f_{\mu\nu}\,,\qquad
\beta_n =\left(\frac{m_f}{m_g}\right)^n\,\tilde\beta_n\,,\qquad
m^4= m^4_g\,\frac{\tilde m^2}{\meff^2}\,.
\ee
In the new variables, the action is invariant under the interchanges
$\tilde g\leftrightarrow\tilde f,\,\tilde\beta_n\leftrightarrow
\tilde\beta_{4-n}$, 
\be 
S_{gf}=\meff^2\int\td^4x \left[\sqrt{-\tilde g}\,R(\tilde g) +
\sqrt{-\tilde f}\,R(\tilde f)-2\tilde m^2\sqrt{-\tilde g}\,
V\left(\sqrt{\tilde g^{-1}\tilde f}\,,\tilde\beta_n\right)\right].
\label{A_action_scaled} 
\ee
The analysis of consistency of coupling the nonlinear massless mode
to matter in section \ref{G-ghost} is performed in terms of these
variables. 

\section{Details of the nonlinear  \texorpdfstring{$G-M^G$}{G-M^G}
  action}  
\label{app_nonlinaction}

Below we work with the rescaled variables introduced above.
For the sake of completeness, here we provide the details for writing
the bimetric action in terms of the nonlinear massless and massive
modes $G$ and $M^G$ (\ref{Gnl}), (\ref{MGnl}). To do this
systematically, note that the expressions for $G$ and $M^G$ can be
inverted to give, 
\be\label{app_gGPhi}
\gmn=G_{\mu\alpha}(\Phi^{-1})^\alpha_{\ph\alpha\nu}\equiv\phi_{\mu\nu}
\,,\qquad
\fmn=G_{\mu\alpha}(\tilde\Phi^{-1})^\alpha_{\ph\alpha\nu}\equiv
\tilde\phi_{\mu\nu}\,. 
\ee
Here $\Phi^\alpha_{\ph\alpha\nu}$ and $\tilde\Phi^\alpha_{\ph\alpha\nu}$
are functions of the matrix $S$ given by,
\be
\Phi^\mu_{\ph\mu\nu} = \delta^\mu_{\nu}
+S^\mu_{\ph\mu\alpha}S^\alpha_{\ph\alpha\nu}\,,\qquad
\tilde\Phi^\mu_{\ph\mu\nu} = \delta^\mu_{\nu}
+(S^{-1})^\mu_{\ph\mu\alpha}(S^{-1})^\alpha_{\ph\alpha\nu}\,,
\label{PhiS}
\ee
and $S$ is related to the massive mode $M^G$ in a simple way,
\be
S=G^{-1}\,M^G +c{\mathbb 1}\,.
\label{SMG}
\ee
These express $g$ and $f$ in terms of $G$ and $M^G$. We also define
the inverse matrices, 
\be\label{app_PhiPhiinv}
\phi^{\mu\alpha}\phi_{\alpha\nu}=\delta^\mu_{\nu}\,,\quad
\tilde\phi^{\mu\alpha}\tilde\phi_{\alpha\nu}=\delta^\mu_{\nu}\,.
\ee
Using (\ref{RNabla}) the curvatures of $g$ and $f$ can be related to
the curvature of $G$ and quantities that contain covariant derivatives
only with respect to $\Gmn$, 
\be\label{RgG}
R_{\mu\nu}(g) = R_{\mu\nu}(G) 
+ 2\nabla_{[\mu}C_{\alpha]\nu}^{\ph\alpha\ph]\ph\nu\alpha}
-2C_{\nu[\mu}^{\ph\nu\ph[\ph\mu\beta}C_{\alpha]\beta}^{\ph\alpha\ph]\ph\beta\alpha}\,,
\ee
where,
\be\label{CgG}
C_{\mu\nu}^{\ph\mu\ph\nu\alpha}=\tfrac1{2}g^{\alpha\beta}\left(
\nabla_\mu g_{\beta\nu}+\nabla_\nu g_{\beta\mu}
-\nabla_\beta \gmn\right)\,.
\ee
Similarly we have that,
\be\label{RfG}
R_{\mu\nu}(f) = R_{\mu\nu}(G) 
+ 2\nabla_{[\mu}\widetilde C_{\alpha]\nu}^{\ph\alpha\ph]\ph\nu\alpha}
-2\widetilde C_{\nu[\mu}^{\ph\nu\ph[\ph\mu\beta}
\widetilde C_{\alpha]\beta}^{\ph\alpha\ph]\ph\beta\alpha}\,,
\ee
where,
\be\label{CfG}
\widetilde C_{\mu\nu}^{\ph\mu\ph\nu\alpha}=\tfrac1{2}f^{\alpha\beta}\left(
\nabla_\mu f_{\beta\nu}+\nabla_\nu f_{\beta\mu}
-\nabla_\beta \fmn\right)\,.
\ee
We further note that the volume densities can be expressed as,
\be\label{app_sqrtgG}
\sqrt{-\det g} = \sqrt{-\det
  G}~\sqrt{\det\Phi^{-1}}\,,\quad\mathrm{and}
\quad
\sqrt{-\det f} = \sqrt{-\det G}~\sqrt{\det\tilde\Phi^{-1}}\,.
\ee
Using these relations we proceed to obtain the general structure of
the nonlinear action. 

{\bf Kinetic structure for $M^G$}:
Using (\ref{app_gGPhi}), (\ref{app_PhiPhiinv}), and
(\ref{app_sqrtgG}), it is a straightforward but tedious algebraic
exercise to show that (modulo a total derivative), 
\begin{align}\label{app_CCterms}
&\sqrt{-\det g}~\phi^{\mu\nu}\left(
2\nabla_{[\mu}C_{\alpha]\nu}^{\ph\alpha\ph]\ph\nu\alpha}
-2C_{\nu[\mu}^{\ph\nu\ph[\ph\mu\beta}
C_{\alpha]\beta}^{\ph\alpha\ph]\ph\beta\alpha}\right)\nn\\[.1cm]
	&\hspace{.5cm}=\sqrt{-\det g}\,
\frac1{4}\biggl(\phi^{\alpha\beta}
\phi_{\kappa\lambda}\phi_{\rho\sigma}
+2\delta^\beta_{\lambda}\delta^\alpha_{\sigma}
\phi_{\kappa\rho}
 -2\delta^\alpha_{\lambda}\delta^\beta_{\rho}
\phi_{\kappa\sigma}
-\phi^{\alpha\beta}\phi_{\kappa\rho}\phi_{\lambda\sigma}
\biggr)\nabla_\alpha\phi^{\kappa\rho}\nabla_\beta\phi^{\lambda\sigma}\,.
\end{align}
A similar result is obtained for the corresponding term in the $\fmn$
sector by simply replacing $\Phi$ by $\tilde\Phi$ everywhere in the
above (and $g$ by $f$ in the determinant prefactor). Next note that
from (\ref{PhiS}) we have,
\be\label{DPhiM}
\nabla_\alpha\Phi^\mu_{\ph\mu\nu}
=\left(\delta^\mu_{\lambda}S^\sigma_{\ph\sigma\nu}
+\delta^\sigma_{\nu}S^\mu_{\ph\mu\lambda}
\right)\nabla_\alpha S^\lambda_{\ph\lambda\sigma}\,,
\ee
and also
\be\label{DPsiM}
\nabla_\alpha\tilde\Phi^\mu_{\ph\mu\nu}
=-\left((S^{-1})^\mu_{\ph\mu\lambda}(S^{-1})^\sigma_{\ph\sigma\beta}
(S^{-1})^\beta_{\ph\beta\nu}+(S^{-1})^\sigma_{\ph\sigma\nu}
(S^{-1})^\mu_{\ph\mu\beta}(S^{-1})^\beta_{\ph\beta\lambda}
\right)\nabla_\alpha S^\lambda_{\ph\lambda\sigma}\,.
\ee
These together with (\ref{app_CCterms}) and its corresponding
expression for $\fmn$ give,
\begin{align}
\label{app_CCgterms2}
\sqrt{-\det g~}\phi^{\mu\nu}\left(
2\nabla_{[\mu}C_{\alpha]\nu}^{\ph\alpha\ph]\ph\nu\alpha}
-2C_{\nu[\mu}^{\ph\nu\ph[\ph\mu\beta}
C_{\alpha]\beta}^{\ph\alpha\ph]\ph\beta\alpha}\right)
&=\sqrt{-\det g}~\mathcal{P}_{\alpha\beta}^{\lambda\sigma
\kappa\rho}~\nabla_\lambda S^\alpha_{\ph\alpha\sigma}
\nabla_\kappa S^\beta_{\ph\beta\rho}\,,
\\
\label{app_CCfterms2}
\sqrt{-\det f}~\tilde\phi^{\mu\nu}\left(2\nabla_{[\mu}\widetilde
C_{\alpha]\nu}^{\ph\alpha\ph]\ph\nu\alpha}
-2\widetilde C_{\nu[\mu}^{\ph\nu\ph[\ph\mu\beta}
\widetilde C_{\alpha]\beta}^{\ph\alpha\ph]\ph\beta\alpha}\right)
&=\sqrt{-\det f}~\widetilde{
\mathcal{P}}_{\alpha\beta}^{\lambda\sigma\kappa\rho}~
\nabla_\lambda S^\alpha_{\ph\alpha\sigma}\nabla_\kappa
S^\beta_{\ph\beta\rho}\,, 
\end{align}
in terms of the polarization tensors $\mathcal{P}$ and
$\widetilde{\mathcal{P}}$. We refrain from writing out the full
expressions for these tensors here and simply note that they can be
straightforwardly deduced from (\ref{app_CCterms}), (\ref{DPhiM}) and
(\ref{DPsiM}). Since,
\be
\nabla_\kappa S^\beta_{\ph\beta\rho}=G^{\beta\lambda}\nabla_\kappa
M^G_{\lambda\rho}\,,
\ee
equations (\ref{app_CCgterms2}) and (\ref{app_CCfterms2}) provide the
kinetic term for massive field $M^G$ with only $G$-covariant
derivatives.  

{\bf Kinetic structure for $G$}:
From (\ref{RgG}) and (\ref{RfG}) combined with (\ref{app_gGPhi}) we 
find the corresponding relations for the Ricci scalars, 
\begin{align}
R(g)&= R(G) + S^\mu_{\ph\mu\alpha}S^\alpha_{\ph\alpha\beta}
G^{\beta\nu}R_{\mu\nu}(G) + \dots \\[.2cm]
R(f)&=R(G)+(S^{-1})^\mu_{\ph\mu\alpha}(S^{-1})^\alpha_{\ph\alpha\beta}  
G^{\beta\nu}R_{\mu\nu}(G) + \dots
\end{align}
where the dots represents the kinetic terms for $M^G$ discussed above.
Hence, the kinetic structure for $G$ is given by the usual
Einstein-Hilbert term plus non-minimal coupling of the Ricci 
tensor to $S$ in the $\gmn$ sector and to $S^{-1}$ in the $\fmn$
sector. Apart from this we also take into consideration the volume
densities given by (\ref{app_sqrtgG}). Thus, the full kinetic
structure for $G$ is given by (omitting the overall factor of
$\sqrt{-\det G}$), 
\begin{align}\label{RofGM}
&\left(\sqrt{\det\Phi^{-1}}+\sqrt{\det\tilde\Phi^{-1}}
\right)~R(G) +\sqrt{\det\Phi^{-1}}~
 S^\mu_{\ph\mu\alpha}S^\alpha_{\ph\alpha\beta}
G^{\beta\nu}R_{\mu\nu}(G)\nn\\[.1cm]
&\hspace{5.3cm} +\sqrt{\det\tilde\Phi^{-1}}
 (S^{-1})^\mu_{\ph\mu\alpha}(S^{-1})^\alpha_{\ph\alpha\beta}~
G^{\beta\nu}R_{\mu\nu}(G)\,.
\end{align}
This relation together with (\ref{app_CCgterms2}) and
(\ref{app_CCfterms2}) completely determine the kinetic terms.   

{\bf Full nonlinear $G-M^G$ action:}
Using (\ref{SMG}), the interaction potential is easily expressed in
terms of $M^G$ as in (\ref{VM}),
\be
\sqrt{-\det g}~V(S,\beta_n) =
\sqrt{-\det G}~(1+S^2)^{-1/2}~V(M^G,\alpha_n^c)\,.
\ee
Collecting all the results, we can now write the Lagrangian in
(\ref{A_action_scaled}) in terms of $G$ and $M^G$ as, 
\begin{align}\label{app_Lagnl}
\Lag(G,M^G)&=\left(\det(1+S^2)^{-1/2}+\det(1+S^{-2})^{-1/2}\right)~R(G)
\nn\\[.1cm]
&\hspace{0.4cm}+
\left(\det(1+S^2)^{-1/2}\mathcal{P}_{\alpha\beta}^{\lambda\sigma
	\kappa\rho}
+\det(1+S^{-2})^{-1/2}\wt{\mathcal{P}}_{\alpha\beta}^{\lambda\sigma
	\kappa\rho}\right)~
	\nabla_\lambda S^\alpha_{\ph\alpha\sigma}
	\nabla_\kappa S^\beta_{\ph\beta\rho}\nn\\[.1cm]
&\hspace{0.4cm} + \det(1+S^2)^{-1/2}~S^\mu_{\ph\mu\alpha}
S^\alpha_{\ph\alpha\beta}G^{\beta\nu}R_{\mu\nu}(G)\nn\\[.1cm]
&\hspace{0.4cm} + \det(1+S^{-2})^{-1/2}(S^{-1})^\mu_{\ph\mu\alpha}
(S^{-1})^\alpha_{\ph\alpha\beta}G^{\beta\nu}R_{\mu\nu}(G)\nn\\[.1cm]
&\hspace{0.4cm} -2m^2\det(1+S^2)^{-1/2}~ V(M^G,\alpha_n^c)\,,
\end{align}
such that the full action is given by
\be
S_{GM} = \meff^2\int\td^4x\sqrt{-\det G}~\Lag(G,M^G)\,.
\ee

\end{document}